\documentclass[aps,prl,twocolumn]{revtex4-1}

\usepackage{graphicx}
\usepackage{epstopdf}
\usepackage{dcolumn}
\usepackage{bm}
\usepackage{amsmath}
\usepackage{amssymb}
\usepackage[dvipsnames]{xcolor}
\usepackage{color} 
\usepackage{ulem}
\usepackage{xspace}

\renewcommand{\tilde}[1]{\widetilde{#1}}

\usepackage{here}

\begin{document}

\title{Quantum Monte Carlo Simulation of Frustrated Kondo Lattice Models}

\author{Toshihiro Sato$^1$}
\author{Fakher F. Assaad$^1$}
\author{Tarun Grover$^{2}$}

\affiliation{
\mbox{$^1$Institut f\"ur Theoretische Physik und Astrophysik, Universit\"at W\"urzburg, 97074 W\"urzburg, Germany}\\
\mbox{$^2$Department of Physics, University of California at San Diego, La Jolla, CA 92093, USA}\\
}

\begin{abstract}
The absence of negative sign problem  in quantum Monte Carlo  simulations of spin and fermion systems has different origins.  World-line based   algorithms for spins require positivity of  matrix elements whereas   auxiliary field approaches for fermions  depend on symmetries such as  particle-hole.   For negative-sign-free spin and fermionic  systems, we show that one can formulate a  negative-sign-free auxiliary field quantum Monte Carlo algorithm that allows Kondo coupling of fermions  with the  spins.  
 Using this  general approach, we study a half-filled  Kondo lattice model on the Honeycomb lattice   with geometric frustration.  In addition to the conventional Kondo insulator and anti-ferromagnetically ordered phases, we find a partial Kondo screened state where spins are  selectively screened so as to alleviate frustration, and the lattice rotation symmetry is broken nematically. 
\end{abstract}

\maketitle

{\it Introduction.}---
Unconventional, highly entangled states can arise if one starts from a system with a large, perhaps infinite, ground state degeneracy, and then perturb it slightly to lift the degeneracy. Fractional quantum Hall systems clearly fall in this category - at any fractional filling the non-interacting problem of electrons in Landau levels has an infinite number of ground states in the thermodynamic limit. Perturbing this system with interactions leads to a particular superposition of these ground states that corresponds to fractional quantum Hall states. Geometrically frustrated spin-systems provide a different class of similar phenomenon.  As an example, consider a square lattice where each link $ij$ that connects vertices $i,j$ hosts a spin-1/2 spin ${\hat{\mathbf{S}}}_{i,j}$ which interact via the Hamiltonian $\hat{H}_{\rm classical} = J \sum_{i,j,k,l \in \Box} \hat{S}^z_{ij} \hat{S}^z_{jk} \hat{S}^z_{kl} \hat{S}^z_{li}$. This  model has an extensive ground state entropy. Now consider a perturbed model: $\hat{H}_{\rm quantum} =  \hat{H}_{\rm classical} + \epsilon \hat{S}^x_i$. For non-zero $\epsilon \ll 1 $, the ground state of this new model is identical to that of  Kitaev's celebrated Toric Code \cite{Kitaev03}: it corresponds to an equal weight superposition of the ground states of  $\hat{H}_{\rm classical}$. Motivated by these examples, we ask: what  phases emerge when a geometrically frustrated spin system is coupled to \textit{fermions}?  In this paper we will describe a quantum Monte Carlo (QMC) algorithm that allows one to study a large class of frustrated magnets Kondo coupled to fermions, and demonstrate the algorithm by studying a specific model that exhibits  new partial Kondo screened (PKS)  phases. 

For concreteness, consider  the following  Hamiltonian of interacting fermions and spins, $\hat{H}= \hat{H}_{\text{Spin}} + \hat{H}_{\text{Fermion}} +  \hat{H}_{\text{Kondo}}$ where
\begin{equation}
\label{eq:frustkondo}
 \hat{H}_{\text{Spin}}   =  \sum_{i,j} \left(  J^z_{ij} \hat{S}^z_{i} \hat{S}^z_{j} + J^{\perp}_{ij} \left(\hat{S}^+_{i} \hat{S}^-_{j} + h.c.\right) \right)  
\end{equation}
\begin{equation*}
\hat{H}_{\text{Fermion}}  = \sum_{x,y,\sigma}   \hat{c}^{\dagger}_{x\sigma} T^{\phantom\dagger} _{x,y} \hat{c}^{\phantom\dagger} _{y\sigma} + \sum_x U (\hat{n}^{\phantom\dagger} _{x,\downarrow}-\frac{1}{2}) (\hat{n}^{\phantom\dagger} _{x,\uparrow}-\frac{1}{2}) 
\end{equation*}
\begin{equation*}
  \hat{H}_{\text{Kondo}}  =  \sum_{i,x}   \frac{J^{\rm K}_{i,x}}{2}   \hat{\pmb{c}}^{\dagger}_x \left[ \sigma^{z}  \cdot \hat{S}_i^{z}  - (-1)^{x} \left(  \sigma^{+}  \hat{S}_i^{-} +  \sigma^{-}  \hat{S}_i^{+} \right)  \right] \hat{\mathbf{c}}^{\phantom\dagger}_x \nonumber
\end{equation*}

Here the spin 1/2 local moments (electrons)  ${\hat{\mathbf{S}}_i}$   $ \left( \hat{\pmb{c}}^{\dagger}_x = \left( \hat{c}^{\dagger}_{x,\uparrow}, \hat{c}^{\dagger}_{x,\downarrow} \right) \right) $    reside on a   graph with sites labeled by $i,j$ ($x,y$).  $J^z_{ij}, J^{\perp}_{ij}$  defines the  potentially frustrated spin model and  $T_{x,y}$  the hopping matrix elements  of conduction electrons subject to electron correlations modeled by a Hubbard U-term  \cite{Hubbard63}.   The local moments and electrons interact via the Kondo coupling $J^{\text{K}}_{i,x}$.   For the sake of generality we have included the phase factor $(-1)^x$ in the Kondo coupling. This phase factor plays no role if the transverse spin interaction is bipartite, or  if the Kondo coupling  includes conduction electron only on one  sublattice. 

It is natural to ask when such Hamiltonians do not suffer from the `sign problem' \cite{Sandvik99b1,Loh1990}, which can make it impossible to simulate quantum systems using finite resources \cite{Troyer05}.  
There are two potential sources of the sign problem: the fermions and the geometrical frustration of spins. Conventionally, these difficulties are tackled in two very different ways. If the fermions were at half-filling on a bipartite lattice, then one can employ a determinantal QMC approach to solve this problem \cite{Blankenbecler81,White89,Loh1990,Assaad08_rev}, whereas for spins, if the condition $ J^{\perp}_{ij} < 0$ is met (which still allows for geometrical frustration \cite{Moessner01, Isakov11}), then one can  employ a worldline QMC or stochastic series expansion (SSE) \cite{Sandvik99b1}. Therefore, it is not obvious how one  should  approach this problem in the presence of the Kondo coupling $J^{\text{K}}$ between the fermions and spins. So far all published studies of frustrated Kondo lattice systems have been limited to non-exact methods, such as dynamical mean-field theory (DMFT) \cite{Aulbach2015}, slave-particle mean-field theory \cite{Pixley2015,Pixley2016}, large-$N$ methods \cite{Coleman10} and variational Monte Carlo (VMC)\cite{Motome2010}. There have also been studies where spins are treated classically \cite{Ishizuka2013}, and which therefore do not capture the physics of the Kondo screening ({\it i.e.\/}, EPR singlet formation between spins and electrons), which is an inherently quantum phenomena. Finally, there has also been progress in simulating a class of models of  fermions interacting with geometrically frustrated quantum spins  \cite{Schattner15,liu_frust17,grover_assaad16,gazit2016}. However, the corresponding algorithm is restricted to spin density-density interactions between local moments and electrons, and does not allow for Kondo coupling between spins and fermions.

 In this paper, we will develop an algorithm to solve Hamiltonians of the form in Eq.~(\ref{eq:frustkondo}) using QMC when $\hat{H}_{\textrm{spin}} $ and $\hat{H}_{\textrm{fermion}}$ are each sign problem-free within bosonic (i.e. $ J^{\perp}_{ij} < 0$ )  and fermionic QMC (i.e. $T_{x,y}$ defines  a bipartite graph),  respectively.  The main innovation is the reformulation of the bosonic problem as a fermionic one by writing spins in terms of Abrikosov fermions \cite{Abrikosov1965electron}: $\hat{\mathbf{S}} = \frac{1}{2}\hat{\mathbf{f}}^{\dagger} {\boldsymbol \sigma} \hat{\mathbf{f}}$, where 
 $\hat{\mathbf{f}}^{\dagger} = \left( \hat{f}^{\dagger}_{\uparrow},  \hat{f}^{\dagger}_{\downarrow}\right)$  is a two-component fermion with the constraint $ \hat{\mathbf{f}}^{\dagger} \hat{\mathbf{f}}^{\phantom\dagger}  = 1$. The constraint is implemented \textit{exactly} by adding Hubbard term $ U_f (\hat{f}_{\uparrow}^{\dagger} \hat{f}^{\phantom\dagger} _{\uparrow} -\frac{1}{2}) (\hat{f}_{\downarrow}^{\dagger} \hat{f}^{\phantom\dagger}_{\downarrow} -\frac{1}{2})$, and taking the limit $U_f \rightarrow \infty$.   Most importantly,  the total $\hat{H}$, including the Kondo coupling $\hat{H}_{\rm Kondo}$, does not have a sign problem either. This is a consequence of the existence of an anti-unitary symmetry under which the Hamiltonian $\hat{H}$ is invariant  \cite{Wu04}.   The demonstration of the absence of the sign problem builds on   Ref.~\cite{Assaad99a,Capponi00}  and is detailed in the Supplemental  Material (SM).
 
\begin{figure}
\centering
\centerline{\includegraphics[width=0.3\textwidth]{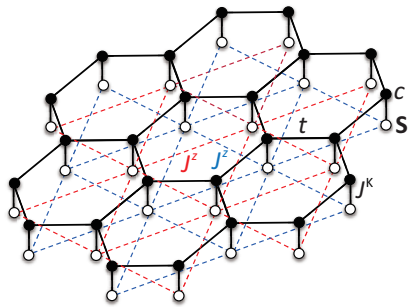}
}
\caption{\label{fig:model}
Kondo lattice model.  The conduction electrons ($c$)   hop  with matrix element $t$ between nearest neighbor sites  of a Honeycomb lattice and are  Kondo coupled  ($J^{\text{K}}$)   to local moments
($\bf{S}$).   Local moments are subject  to geometrical frustration generated by the  next-nearest-neighbor antiferromagnetic Ising interaction $J^{\rm z}$.
}
\end{figure}

The relevance of this class of models to heavy fermion phenomenology alluded above is worth elaborating upon. A simple picture to capture the global phase diagram of heavy fermions was provided by Doniach \cite{Doniach77}. For a single impurity Kondo problem, the cross-over  to the spin-singlet state takes place at the Kondo temperature  $ T_{\rm K} = D \,e^{-1/(2N(E_{\rm F})J^{\text{K}} )}$ where $N(E_{\rm F})$ is the conduction electrons density  of states at the Fermi level $E_{\rm F}$, $J^{\text{K}}$ is the exchange interaction between the localized impurity and the conduction electrons, and $D$ is the conduction electrons bandwidth \cite{Hewson}. Now consider a dilute matrix of such local moments. The conduction electrons will mediate long-range RKKY exchange interaction between the local moments whose scale is given by the temperature $T_{\rm RKKY} \propto (J^{\text {K} })^2N(E_{\rm F})$.  When $T_{\rm K} \gg T_{\rm RKKY}$, one obtains the heavy fermion liquid state, which is the many-body analog of the single impurity's spin-singlet ground state. In contrast, in the opposite limit, the spins are likely to order resulting in an antiferromagnetic metal. However, as hinted above, there is a growing list of materials such as CePdAl, Pr${}_2$Ir${}_2$O${}_7$, YbAgGe, YbAl${}_3$C${}_3$, Yb${}_2$Pt${}_2$Pb \cite{akito16,nakatsuji06, kim_frustkondo08,sengupta_frustkondo10,kato_frustkondo08}, where one observes phases which do not easily fit into either of the two limits Doniach considered. The microscopies of these materials suggests that geometrical frustration plays a crucial role in their phenomenology. Therefore, one is motivated to consider a phase diagram where geometrical frustration is an axis in addition to the Kondo coupling.

\begin{figure}
\centering
\centerline{\includegraphics[width=0.42\textwidth]{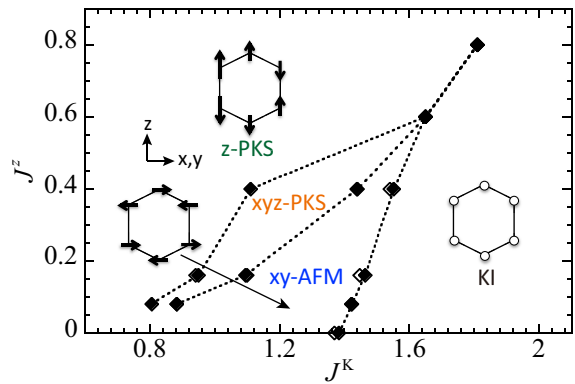}
}
\caption{\label{fig:phasediagram}
Phase diagram with in-plane antiferromagnetic (xy-AFM), out-of-plane partial Kondo screening (z-PKS), spin-rotation symmetry breaking partial Kondo screening (xyz-PKS), and Kondo insulator (KI) phases from QMC simulations at $T=0.025$.
Inset: schematic local moment structure in each phase.
Diamonds indicate the onset of long-range order; Filled (open) symbols are critical values
based on $L= 6$ and 9 ($L=9$ and 12), see text. 
}
\end{figure}

{\it Case study }---
For concreteness, we consider the following model  (see Fig.~\ref{fig:model}): 
\begin{eqnarray}
& & \hat{H}_{\text{Spin}} = J^{z}\sum_{\langle \langle i,j \rangle \rangle}\hat{S}_{i}^{z}\hat{S}_{j}^{z},  \; \;  \hat{H}_{\text{Fermion}} = -t\sum_{\langle i,j \rangle,\sigma}\hat{c}_{i\sigma}^\dagger \hat{c}^{\phantom\dagger} _{j\sigma}  \nonumber \\ 
 & & \hat{H}_{\text{Kondo}}  =  
  J^{\text{K}} \sum_{i}    \frac{1}{2} \hat{\pmb{c}}^{\dagger}_{i} \pmb{\sigma}\hat{\pmb{c}}^{\phantom\dagger}_{i} \cdot \hat{{\bm S}}^{\phantom\dagger} _{i} 
\end{eqnarray}
In this special case  $J^{\perp}_{i,j} = 0$,  and the spins and conduction electrons reside  on the  same Honeycomb lattice so that  we can use the same indices from spins and conduction electrons. Furthermore,  the canonical transformation $ \hat{S}^{\pm}_{i} \rightarrow  - (-1)^{i} \hat{S}^{\pm}_{i }, 
 \hat{S}^{z}_{i} \rightarrow   \hat{S}^{z}_{i } $ will remove the factor $(-1)^{i}$ in the Kondo  coupling of  Eq.~(\ref{eq:frustkondo}).
While $  \hat{H}_{\text{Fermion}} $ and $\hat{H}_{\text{Kondo}} $ account for the generic Kondo lattice model on the Honeycomb lattice, 
$ \hat{H}_{\text{Spin}}  $   is geometrically frustrating since it couples  antiferromagnetically local moments on the two underlying triangular Bravais lattices  of the Honeycomb graph. While this term breaks down the SU(2) total spin symmetry to U(1), time reversal symmetry, essential for the  Kondo  effect, is present.  

For the numerical simulations we used the ALF (Algorithms for Lattice Fermions)
implementation \cite{ALF_v1} of the well-established finite-temperature auxiliary-field QMC method \cite{Blankenbecler81,Assaad08_rev}.  In the SM, it is shown how to rewrite the model such that it will  comply to the data structure of the ALF \cite{ALF_v1}.
We simulated lattices with $L\times L$ unit cells (each containing four orbitals) and periodic boundary conditions.
Henceforth, we use $t=1$ as the energy unit and consider half-filling for the conduction electron.
All the data are calculated for temperature $T=0.025$ (with Trotter discretization $\Delta\tau=0.1$). In the considered parameter range this choice of temperature is representative of the ground state. 

\begin{figure}
\centering
\centerline{\includegraphics[width=0.42\textwidth]{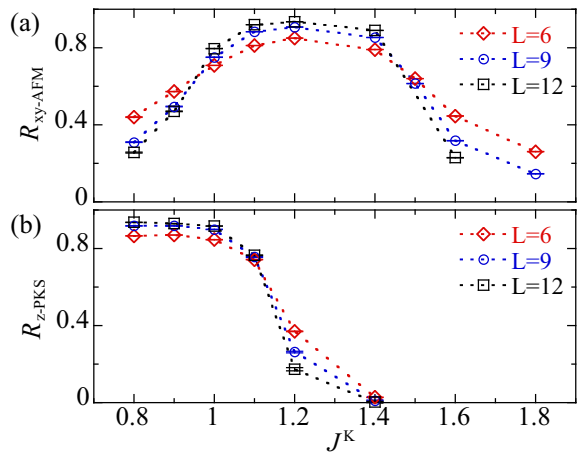}}
\caption{\label{fig:R-Iz0.04}
$J^{\text{K}}$ dependence of correlation ratios for (a) in-plane antiferromagnetic and (b) out-plane three-sublattice orders.
Here, $J^{z}=0.16$ and $T=0.025$.
}
\end{figure}

{\it Phase diagram.}---
Fig.~\ref{fig:phasediagram} shows the phase diagram in the Kondo,  $J^{\text{K}}$, versus  frustration, $J^{z}$, plane 
as obtained from  a finite-size scaling analysis.  To map out the magnetic phase diagram we compute  correlation functions of the total spin, 
 $C^{\alpha}(\mathbf k)\equiv\frac{1}{V}\sum_{\mathbf r, \mathbf r'}\langle  \hat{O}^{\alpha}_{\mathbf r} \hat{O}_{\mathbf r'}^{\alpha}\rangle e^{i \mathbf k (\mathbf r-\mathbf r')}$  where
  $  \hat{O}_{\mathbf r}^{\alpha} = \hat{S}^{tot,\alpha}_{\mathbf{ r},A} - \hat{S}^{tot,\alpha}_{\mathbf{r},B} $  and
 $\hat{S}^{tot,\alpha}_{i}= \frac{1}{2} \hat{c}^{\dagger}_{i} \sigma^{\alpha} \hat{c}^{\phantom \dagger}_{i} +\hat{S}_{i}^{\alpha}$ with $\alpha=(x,y,z)$.
 Here ${\mathbf{ r}} $ labels the unit cell of the Honeycomb lattice and $A$, $B$ the orbitals  \footnote{We have checked that no instabilities occur in  the 
 ferrromagnetic channel:  $  \hat{O}_{\mathbf r}^{\alpha} = \hat{S}^{tot,\alpha}_{\mathbf{ r},A} + \hat{S}^{tot,\alpha}_{\mathbf{r},B} $}. 

We find four phases in the range of parameters showed in Fig.\ref{fig:phasediagram}. The phase diagram along $J^z=0$ axis has been studied earlier \cite{Assaad99a,Capponi00}, and reflects the aforementioned competition between RKKY and Kondo screening with an antiferromagnetic (AFM) phase at small $J^{\text{K}}$, and a Kondo insulator (KI) at large $J^{\text{K}}$. At $J^z$ precisely equal to zero, the model has an SU(2) symmetry and therefore the AFM order parameter can point in along any direction in the spin-space. At infinitesimally small non-zero value of $J^z$, the spins preferentially order in the x-y plane to minimize the energy cost of geometrical frustration. Hence this phase is characterized by   diverging $C^{x/y}(\bm k = \Gamma)$ and  we denote it as xy-AFM in Fig.\ref{fig:phasediagram}.  As the geometrical frustration is increased,  the phase diagram changes dramatically. We find two new phases which we denote as  z-PKS and xyz-PKS where the acronym PKS stands for partially Kondo screened. In the z-PKS phase, the U(1) spin-rotation symmetry is unbroken while the time reversal symmetry corresponding to the operation  $\hat{S}^{tot,z}_{i} \rightarrow  -\hat{S}^{tot,z}_{i}$  is broken. Therefore, this phase is characterized by 
a diverging $ C^{z}(\bm k = \mathbf{K})$  where $\mathbf{K}$   corresponds the  Dirac points of the tight   binding conduction electron model. 
Thereby the z-PKS phase has a  $\sqrt{3}\times\sqrt{3}$ unit cell  depicted in the inset of Fig.~\ref{fig:phasediagram}. The existence of Kondo screening is crucial to understand the qualitative features of the z-PKS phase, as discussed in detail below. The xyz-PKS phase is a canted version of z-PKS and can be thought of as a  hybrid between xy-AFM and z-PKS in that it breaks the symmetries that are broken in either of these phases. 

To locate the phase boundaries  we consider the renormalization group invariant quantity~\cite{Binder1981,Pujari16} 
\begin{equation}
R_{\alpha}=1-\frac{C^{\alpha}(\bm k_0+\delta {\bm  k}) }{C^{\alpha}(\bm k_0)}.
\label{eq:HI}
\end{equation}
Here ${\bm k}_0$ is the ordering wave vector and $ \delta {\bm k}$  the smallest  wave vector on the lattice.   
By definition, $R_{\alpha} \to 1$ for $L\to\infty$ in the  ordered state whereas $R_{\alpha} \to 0$ in the disordered phase.
At the critical point, $R_{\alpha}$ is scale-invariant for sufficiently large $L$ so that results for different system sizes cross.
Figures~\ref{fig:R-Iz0.04}(a) and (b) show typical results at  $J^{z}=0.16$.
The phase boundaries in Fig.~\ref{fig:phasediagram} are based on the crossing points of results for $L=6,9$ (filled symbols) and $L=9,12$ (open symbols),
respectively. 

\begin{figure}
\centering
\centerline{\includegraphics[width=0.42\textwidth]{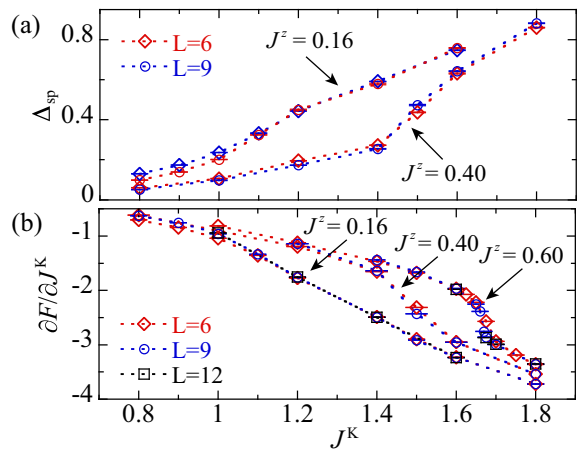}}
\caption{\label{fig:dHdJ}
(a) Single-particle gap $\Delta_\text{sp}$ at the Dirac point and (b) free-energy derivative $\partial F/\partial J^\text{K}$. Here, $T=0.025$. The gap is  obtained from the asymptotic behavior of the imaginary time single particle  Green function \cite{Assaad96a}.   
}
\end{figure}

{\it The z-PKS phase.}---
The  atomic limit ($t=0$) reveals aspects of the z-PKS phase.  Here  the A and B sublattices decouple  to form two independent triangular  lattices.   Translation symmetry breaking of the z-PKS phase  leads to a unit cell, $R$, for a single sublattice, consisting of three distinct  sites, $n$, each  accommodating a spin and  a conduction electron.  A simple  variational ansatz for the wave function is the product state  
$|\Psi_0 \rangle   = \prod_{R,n} ( \alpha_{n,0}  |0,0 \rangle_{R,n}    + \sum\limits_{\mu = -1,0,1} \alpha_{n,\mu} |1,\mu \rangle_{R,n}) $  where 
$|0,0 \rangle_{R,n}  $, $|1,\mu \rangle_{R,n} $  denote  singlet and triplet states of the spin and conduction electrons. The normalization   condition 
$|\boldsymbol{\alpha}_n| = 1 $ holds.    The  variational energy   per unit cell  takes the form 
$ E = \sum_{n} \left(  J^{\text{K}} K_n  -  \frac{3}{32}  J^{z} M_n^{2} \right)   +   \frac{3}{32}  J^{z} \left( \sum_{n} M_n \right)^2 $   with 
$K_n = \langle \Psi_0   | \frac{1}{2} \hat{\pmb{c}}^{\dagger}_{R,n} \pmb{\sigma}\hat{\pmb{c}}^{\phantom\dagger}_{R,n} \cdot \hat{{\bm S}}^{\phantom\dagger} _{R,n} | \Psi_0 \rangle$ and 
$M_n = \langle \Psi_0  | \hat{S}^{z}_{R,n} | \Psi_0 \rangle $.  As apparent from this form,  Kondo screening competes with  the geometric frustration  \footnote{Due to the normalization condition,  $|\boldsymbol{\alpha}_n| = 1 $, finite values of $M_n$  lead to a reduction of the Kondo screening.  }  and it is energetically favorable to set  $\sum_{n} M_n  =0 $.   This condition is by no means  imposed by symmetries and we have thus checked that our realizations of the  z-PKS phase in the  QMC simulations indeed satisfy this condition approximately (see the SM). 

\begin{figure}
\centering
\centerline{\includegraphics[width=0.42\textwidth]{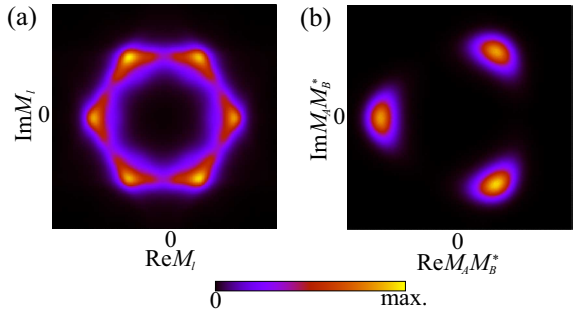}}
\caption{\label{fig:P-Iz0.15}
Probability distribution of (a) $\bm M_l$ and (b) ${\bm M}_A {\bm M}^{*}_B$
for the z-PKS phase at $J^{z}  =  0.60$ and $J^{\text{K}}=1.5$ (see the text for definition).
Here, $T=0.025$ and  $L=9$.}
\end{figure}

The QMC histogram in the complex plane of   
\begin{equation}
{\bm M}_{l}=M_{1l}e^{i0}+M_{2l}e^{i\frac{2\pi}{3}}+M_{3l}e^{i\frac{4\pi}{3}},
\label{eq:HI-1}
\end{equation}
 uniquely  reveals the spin structure.   
Here the additional index $l$ runs over sublattices A and B.  Figure~\ref{fig:P-Iz0.15}(a) plots this quantity, and as detailed in the SM  corresponds to the six-fold degenerate  state
$(M_{1A},M_{2A},M_{3A})=\tilde{m}(2,-1,-1)$ and $(M_{1B},M_{2B},M_{3B})=\tilde{m}(-2,1,1)$. For example, at $J^{z}  =  0.60$ and $J^{\text{K}} = 1.5$, $\tilde{m} = 0.1$.
Away from the atomic limit,  the two sublattices  couple.  The histogram of the quantity ${\bm M}_A {\bm M}^{*}_B$ shown in  Fig.~\ref{fig:P-Iz0.15}(b) demonstrates (see SM) 
that the two sublattices lock in as  depicted in  Fig.~\ref{fig:phasediagram}.    

\textit{Single particle gap.}---    To set the notation, we write the low energy theory of Dirac fermions on the honeycomb lattice as $  \hat{H} _{\text{Dirac}}   =   \sum_{\pmb{p}}  \hat{\Psi}^{\dagger}(\pmb{p})      \left[  p_x \tau^{x}      + p_y \tau^{y} \right] 
\hat{\Psi}(\pmb{p}) $ (see SM for details). The $\boldsymbol{\tau}$ Pauli matrices act on the sublattice index. The spinors $\hat{\Psi}$ also carry a spin-index and a valley index, which are acted upon by the Pauli matrices $\boldsymbol{\sigma}$ and $\boldsymbol{\mu}$ respectively. 

 In the large $J^{\text{K}}$ limit, one obtains a Kondo insulator, whose ground state may be approximated by a direct product of Kondo singlets between  the spin and conduction electron on each site. The single particle gap corresponds to the energy cost of breaking a singlet and is set by $J^{\text{K}}$ \cite{Tsunetsugu97_rev}. At the mean-field level, the xy-AFM magnetic ordering induces  a mass term $M_{x,y} = \langle \hat{\Psi}^{\dagger} \tau^z \sigma_{x,y} \mu^z  \hat{\Psi} \rangle $ of magnitude $J^{\text{K}}$  such that $\Delta_{\text{sp}} \propto J^{\text{K}} $.  This is consistent with the data at  $J^{z}  =  0.16$ shown in Fig.~\ref{fig:dHdJ}.   In contrast, the z-PKS phase retains the U(1) spin rotation symmetry but instead breaks time reversal, lattice translation and point group symmetries.  If  the sum of the magnetic moments in both sub-lattices vanishes  (i.e. $\sum_{m} M_m  =0 $) then the Dirac points will only shift along the x-direction  and no single particle gap opens. This is because in the low energy theory, such an order parameter corresponds to the term  $\hat{\Psi}^{\dagger}(\pmb{p})   \tau^{x}  \sigma^{z}   \hat{\Psi}(\pmb{p})$ which is \textit{not} a Dirac mass since it does not anti-commute with the low energy Hamiltonian. However,  the Kondo screening is still present  in the z-PKS phase as evident by the small value of the magnetic order parameter along the z-direction. Therefore, we expect that the mass scale will  be set by Kondo effect and will depend non-perturbatively on $J^{\text{K}}$ as in the single spin Kondo problem. On the other hand, if the condition $\sum_{m} M_m  =0 $ is not satisfied, a mass term proportional to $J^{\text{K}}$ will be generated in the z-PKS phase. As noted earlier, numerically we find that the condition $\sum_{m} M_m  =0 $ is satisfied to a very good approximation. Such a transition from a perturbative to a non-perturbative mass is in qualitative agreement with Fig.~\ref{fig:dHdJ} where one notices that the single particle gap drops as one enters the PKS phase  when increasing the frustration.  A precise  determination of $\Delta_{\text{sp}} $ in this phase is difficult since nematicity  allows the Dirac points to meander. 

\textit{ Phase transitions.}---
Figure~\ref{fig:dHdJ} plots $\partial F/\partial J^{\text{K}}=\langle \frac{1}{2} \hat{\pmb{c}}^{\dagger}_{i} \pmb{\sigma}\hat{\pmb{c}}^{\phantom\dagger}_{i} \cdot \hat{{\bm S}}^{\phantom\dagger} _{i} \rangle$  along various $J^{z}$ cuts.  
We interpret the absence of  jump in this  quantity in terms of a continuous quantum phase transitions. 
Taking into account  time reversal and translation symmetry breaking,  the z-PKS phase has a  6-fold degeneracy  and  can be described by  an XY model with C$_6$ anisotropy.  C$_6$ anisotropy is irrelevant  at criticality   such that the z-PKS phase can  be characterized in terms of an  effective emergent U(1) symmetry. 
The xy-AFM phase is characterized by broken U(1) spin  symmetry.  In the phase diagram of Fig~\ref{fig:phasediagram}   all phase translation lines are characterized by the  spontaneous symmetry breaking of only one of  the two aforementioned U(1) symmetries. Thereby   we expect all quantum phase transitions to  belong to the (2+1)D XY universality class.

{\it Summary and discussion.}--- Using a  fermion representation of the spin-1/2  algebra, we have introduced a large class of Kondo lattice models  (see Eq.~(\ref{eq:frustkondo})) that are free of the negative sign  problem within the auxiliary field QMC approach. Essentially we require the spin system to be free of sign problem in world-line type approaches  and the fermionic system to be particle-hole symmetric  such that  auxiliary field approaches are equally sign free.    This insight gives the possibility of tackling a number of Kondo lattice problems  where frustration  plays a central role in understanding the phase diagram.  It is of experimental relevance since geometrical frustration  is present in many heavy fermion materials \cite{akito16,nakatsuji06, kim_frustkondo08,sengupta_frustkondo10,kato_frustkondo08}.
 
 We have used  our approach to  compute the phase diagram of the Kondo lattice model  on the  honeycomb lattice with geometrical  frustration thus adding a new axis in the generic Doniach phase  diagram. 
 Aside from the RKKY   driven AF  order  (xy-AFM) with broken U(1) spin symmetry and the Kondo   state  with the full microscopic symmetries of the model,  we observe a  novel phase  (z-PKS)  driven by geometrical frustration.    This phase   has U(1) spin symmetry but breaks time reversal,  lattice  and point group symmetries.  It can be understood as a realization of partial Kondo screening in the sense that the strength of Kondo screening becomes site dependent so as to accommodate  frustration.  As opposed to non-frustrated models 
 \cite{Assaad99a,Tsunetsugu97_rev}, 
 the magnetic ordering in the z-PKS phase,  does not necessarily lead to the opening of a single particle gap.  To the best of our knowledge, this is first realization of this type state  using    approximation free exact methods.  Although our  Hamiltonian is not constructed to  model a specific material, it is worth noting that a distinct feature of geometrically frustrated heavy-fermion materials such as CePdAl \cite{akito16} is that similar to the z-PKS phase, they host magnetically ordered phases where the unit-cell is enlarged and different sites within a unit cell have a different value of the magnetic order parameter.

{\it Acknowledgments.}---
We would like to thank M.  Raczkowski  for fruitful discussions. 
This work was supported by the German Research Foundation (DFG) through SFB 1170 ToCoTronics and FOR~1807. 
We gratefully acknowledge the Gauss Centre for Supercomputing (GCS) for allocation of CPU time on the SuperMUC computer at the Leibniz Supercomputing Center
as well as the John von Neumann Institute for Computing (NIC) for computer resources on the JURECA~\cite{Jureca16} machine at the J\"ulich Supercomputing Centre (JSC).
TG acknowledges support from the UCSD startup funds and is also supported as a Alfred P. Sloan Research Fellow. This work was initiated at the Kavli Institute for Theoretical Physics (KITP) during the program `Entanglement in Strongly-Correlated Quantum Matter' and correspondingly, this research was supported in part by the National Science Foundation under Grant No. NSF PHY-1125915.  We equally acknowledge the Bavaria California Technology Center (BaCaTeC) for financial support.

\section{Supplemental Material}

In this supplemental material section we will first give some details on how to formulate  a  negative sign free QMC simulation  for the general Hamiltonian of Eq.~(\ref{eq:frustkondo})  of the main text. We will then proceed in describing  how one can  extract  the spin-ordering in the z-PKS phase using the method of histograms.  Finally  we show that  the z-PKS spin ordering triggers a nematic transition in the Dirac  spectrum and  that it does not necessarily  open a  mass gap. 

\subsection{Monte Carlo Algorithm}

In this section we detail the formulation of the auxiliary field QMC algorithm for the  model
$H= H_{\text{Spin}} + H_{\text{Fermion}} + H_{\text{Kondo}}$  where
\begin{eqnarray}
& & H_{\text{Spin}}   =  \sum_{i,j} \left(  J^z_{ij} S^z_{i} S^z_{j} + J^{\perp}_{ij} \left(S^+_{i} S^-_{j} + h.c.\right) \right)  \\
& & H_{\text{Fermion}}  = \sum_{x,y,\sigma}   c^{\dagger}_{x\sigma} T^{\phantom\dagger}_{x,y} c_{y\sigma} + U \sum_x  (n^{\phantom\dagger}_{x,\downarrow}-\frac{1}{2}) (n^{\phantom\dagger}_{x,\uparrow}-\frac{1}{2}) \nonumber \\
& &  H_{\text{Kondo}}  =  \sum_{i,x}  J^{\rm K}_{i,x}  \mathbf{c}^{\dagger}_x \left[ \frac{\sigma^{z}}{2}  \cdot S_i^{z}  - \frac{(-1)^{x}}{2} \left(  \sigma^{+}  S_i^{-} +  \sigma^{-}  S_i^{+} \right)  \right] \mathbf{c}^{\phantom\dagger} _x .
\nonumber
 \label{eq:frustkondo1}
\end{eqnarray}
To simplify the notation we  omit \textit{ hats} on second quantized operators.  $\mathbf{c}^{\dagger}_{x}  = \left(   c^{\dagger}_{x\uparrow} ,  {c}^{\dagger}_{x,\downarrow} \right)  $   is a fermionic spinor, 
$n^{\phantom\dagger}_{x,\sigma} = c^{\dagger}_{x,\sigma}  c^{\phantom\dagger}_{x,\sigma} $ and  $\mathbf{S}^{\phantom\dagger}_i$  a spin-1/2 degree of freedom.
We consider two separate graphs, one for the conduction  electrons  ($x,y$-indices)  and one  for the spin ($i,j $-indices) degrees of freedom.  The conditions for the absence of sign problem are 
\begin{itemize}
\item   $J^{\perp}_{i,j} \leq 0 $,  $J^{K}_{i,x} \geq 0$ and $U  \geq 0$.
\item    The  conduction electron graph has a bipartition A, B such that $T_{x,y} \neq 0 $ only if  $x$ and $y$ are in different sublattices.  Given the bipartition, $(-1)^{x} = 1 $ ($-1$)  for 
$x\in A$ ($B$).
\end{itemize}

To formulate the  algorithm  we adopt a fermion representation of the  spin-1/2 degree of freedom
\begin{equation}
	\mathbf{S} = \frac{1}{2} \mathbf{f}^{\dagger}_{i}  \boldsymbol{\sigma}  \mathbf{f}^{\phantom\dagger}_{i}  
\end{equation}
with constraint
\begin{equation}
	 \mathbf{f}^{\dagger}_{i}  \mathbf{f}^{\phantom\dagger}_{i}   = 1. 
	 \end{equation}
Here, 
$  \mathbf{f}^{\dagger}_{i}  \equiv  \left(  
	 f^{\dagger}_{i,\uparrow}, f^{\dagger}_{i,\downarrow} \right) $ 
and $ \boldsymbol{\sigma} $    corresponds to the  vector of Pauli spin-1/2 matrices.
It is convenient to work in the Bogoliubov basis,
\begin{eqnarray}
	& & \tilde{\mathbf{f}}^{\dagger}_{i} \equiv  \left( \tilde{f}^{\dagger}_{i,\uparrow},  \tilde{f}^{\dagger}_{i,\downarrow} \right)   =  \left( f^{\dagger}_{i,\uparrow},  f^{\phantom\dagger}_{i,\downarrow} \right)   \nonumber \\
	& & \tilde{\mathbf{c}}^{\dagger}_{x} \equiv  \left( \tilde{c}^{\dagger}_{i,\uparrow},  \tilde{c}^{\dagger}_{i,\downarrow} \right)   =   \left( c^{\dagger}_{i,\uparrow},  (-1)^{x} c^{\phantom\dagger}_{x,\downarrow} \right),
\end{eqnarray}
and to  note that 
\begin{widetext}
\begin{eqnarray}
       -\frac{1}{4}  \left(  \tilde{\mathbf{f}}^{\dagger}_{i}   \tilde{\mathbf{f}}^{\phantom\dagger}_{j}  +  \tilde{\mathbf{f}}^{\dagger}_{j}   \tilde{\mathbf{f}}^{\phantom\dagger}_{i}   \right)^2   & = &
    S^{z}_{i}S^{z}_{j} - \frac{1}{2} \left(  S^{+}_{i}S^{-}_{j} + S^{-}_{i}S^{+}_{j}  \right)  + \boldsymbol{\eta}_{i} \cdot \boldsymbol{\eta}_{j}  \nonumber  \\
      -  \frac{1}{4}\left( \left( \tilde{\mathbf{f}}^{\dagger}_{i}  \tilde{\mathbf{f}}^{\phantom\dagger}_{i}  -1 \right) \pm \left( \tilde{\mathbf{f}}^{\dagger}_{j}   \tilde{\mathbf{f}}^{\phantom\dagger}_{j} -1 \right) \right)^2   & = &    \mp  2 S^{z}_{i} S^{z}_{j}   - (S^{z}_{i})^{2}  -  (S^{z}_{j})^{2}   \nonumber \\
     -\frac{1}{4}  \left(  \tilde{\mathbf{f}}^{\dagger}_{i}   \tilde{\mathbf{c}}^{\phantom\dagger}_{x}  +  \tilde{\mathbf{c}}^{\dagger}_{x}   \tilde{\mathbf{f}}^{\phantom\dagger}_{i}   \right)^2   & = & 
      {\mathbf c}^{\dagger}_x \left[ \frac{\sigma^{z}}{2}  \cdot S_i^{z}  - \frac{(-1)^{x} }{2} \left(  \sigma^{+}  S_i^{-} +  \sigma^{-}  S_i^{+} \right)  \right] {\mathbf c}_x  +  \nonumber \\
    & &  \frac{1}{2} \left( {\mathbf c}^{\dagger}_{x} {\mathbf c} ^{\phantom\dagger}_{x}  - 1 \right) \eta^{z}_i   + (-1)^x \left( c^{\dagger}_{x,\uparrow}  c^{\dagger}_{x,\downarrow}  \eta^{-}_{i} +   
       c^{\phantom\dagger}_{x,\downarrow}  c^{\phantom\dagger}_{x,\uparrow}   \eta^{+}_{i} \right)  \nonumber \\ 
       - \frac{1}{2}\left( \tilde{\mathbf c}^{\dagger}_x  \tilde{\mathbf c}^{\phantom\dagger}_x -1 \right)^2  & = &   \left( n^{\phantom\dagger}_{x,\uparrow} - \frac{1}{2} \right)   \left( n^{\phantom\dagger}_{x,\downarrow} - \frac{1}{2}  \right) -  \frac{1}{4}    \nonumber \\    
     \tilde{\mathbf c}^{\dagger}_x T_{x,y}   \tilde{\mathbf c}^{\phantom\dagger}_y   & = &     {\mathbf c}^{\dagger}_x T_{x,y}  { \mathbf c}^{\phantom\dagger}_y. 
\end{eqnarray}
\end{widetext}
The last equation  holds since $T_{x,y}$ is bipartite. The $ \boldsymbol{\eta}_i$ operators read 
\begin{equation}
  \eta^z_{i}  =  \frac{1}{2} \left( {\mathbf f}^{\dagger}_{i} {\mathbf f} ^{\phantom\dagger}_{i}  - 1 \right) ,  
  \eta^+_{i}  =  f^{\dagger}_{i,\uparrow}  f^{\dagger}_{i,\downarrow}, 
  \eta^-_{i}   =  f_{i,\downarrow}  f_{i,\uparrow} .
\end{equation}
The Hamiltonian  that we will  simulate  reads: 
\begin{widetext}
\begin{eqnarray}
	 H_{\text{QMC}}  & = &   -\sum_{i,j} \left| J^{\perp}_{i,j} \right|  \left[    \frac{1}{2}  \left(  \tilde{\mathbf{f}}^{\dagger}_{i}   \tilde{\mathbf{f}}^{\phantom\dagger}_{j}  +  \tilde{\mathbf{f}}^{\dagger}_{j}   \tilde{\mathbf{f}}^{\phantom\dagger}_{i}   \right)^2     
	                 + \frac{1}{4}\left( \left( \tilde{\mathbf{f}}^{\dagger}_{i}  \tilde{\mathbf{f}}^{\phantom\dagger}_{i}  -1 \right) + \left( \tilde{\mathbf{f}}^{\dagger}_{j}   \tilde{\mathbf{f}}^{\phantom\dagger}_{j} -1 \right) \right)^2  \right]  \nonumber \\
	 & &  -\frac{1}{8} \sum_{i,j}  | J^{z}_{i,j} |      \left( \left( \tilde{\mathbf{f}}^{\dagger}_{i}  \tilde{\mathbf{f}}^{\phantom\dagger}_{i}  -1 \right) - 
	    \frac{ J^{z}_{i,j} } {| J^{z}_{i,j} | }\left( \tilde{\mathbf{f}}^{\dagger}_{j}   \tilde{\mathbf{f}}^{\phantom\dagger}_{j} -1 \right) \right)^2 
	     -U_f  \sum_{i} \left(  \tilde{\mathbf{f}}^{\dagger}_{i}  \tilde{\mathbf{f}}^{\phantom\dagger}_{i}  -1  \right)^2  \nonumber \\  
	 & &  + \sum_{x,y} \tilde{\mathbf c}^{\dagger}_x T^{\phantom\dagger}_{x,y}   \tilde{\mathbf c}^{\phantom\dagger}_y    -\frac{U}{2}  \sum_{x} \left(  \tilde{\mathbf{c}}^{\dagger}_{x}  \tilde{\mathbf{c}}^{\phantom\dagger}_{x}  -1  \right)^2
	  -\frac{1}{4}  \sum_{i,x} J^{\text{K}}_{i,x} \left(  \tilde{\mathbf{f}}^{\dagger}_{i}   \tilde{\mathbf{c}}^{\phantom\dagger}_{x}  +  \tilde{\mathbf{c}}^{\dagger}_{x}   \tilde{\mathbf{f}}^{\phantom\dagger}_{i}   \right)^2    
\end{eqnarray}
\end{widetext}
It is important to note that 
\begin{equation}
	\left[  \left(  \tilde{\mathbf{f}}^{\dagger}_{i}  \tilde{\mathbf{f}}^{\phantom\dagger}_{i}  -1  \right)^2 , H_{\text{QMC}}   \right]  = 0 
\end{equation}
such that   f-fermion parity  $ 2 \left(  \tilde{\mathbf{f}}^{\dagger}_{i}  \tilde{\mathbf{f}}^{\phantom\dagger}_{i}  -1  \right)^2  - 1 =   8 \left( S^{z}_{i} \right)^{2} - 1 = -(-1)^{ \mathbf{f}^{\dagger}_{i}  \mathbf{f}^{\phantom\dagger}_{i} }  $  is a local conserved quantity.   Due to this symmetry property, terms of the form  $-U_f  \sum_{i} \left(  \tilde{\mathbf{f}}^{\dagger}_{i}  \tilde{\mathbf{f}}^{\phantom\dagger}_{i}  -1  \right)^2$, with $U_f > 0$,  will  project  very efficiently on the $ (-1)^{\mathbf{f}^{\dagger}_{i}  \mathbf{f}^{\phantom\dagger}_{i} } =-1 $ subspace thereby  imposing  the constraint 
$ \mathbf{f}^{\dagger}_{i}  \mathbf{f}^{\phantom\dagger}_{i} = 1 $.     Similar ideas were used in the framework of the Kondo lattice model in the absence of frustration \cite{Assaad99a,Capponi00}.  In this subspace, 
$ \boldsymbol{\eta}_i = 0$ and $ \left(S^{z}_{i} \right)^2 = 1/4 $ such that 
\begin{equation}
\left. H_{\text{QMC}}   \right|_{ (-1)^{\mathbf{f}^{\dagger}_{i}  \mathbf{f}^{\phantom\dagger}_{i} } =1} = H.
\end{equation}

The interaction part of $H_{\text{QMC}} $  is a sum of perfect squares of single body operators   with particle number conservation in the Bogoliubov  basis.  It can  thus be implemented in the ALF   (Algorithms for Lattice Fermions)  \cite{ALF_v1}  implementation of the auxiliary field QMC algorithm \cite{Blankenbecler81,White89,Assaad08_rev}.  The absence of sign problem stems from the fact that the coefficients of the  perfect square terms are all negative  and that the single body operators commute with the anti-unitary  operator $T$, 
\begin{equation}
      T^{-1}  \alpha
\begin{pmatrix}
	 \tilde{c}_{i,\uparrow}  \\
	 \tilde{c}_{i,\downarrow}  \\
	 \tilde{f}_{i,\uparrow}  \\
	 \tilde{f}_{i,\downarrow}  
\end{pmatrix}
T = 
\overline{\alpha} 
\begin{pmatrix}
	- \tilde{c}_{i,\downarrow}  \\
	\phantom{-} \tilde{c}_{i,\uparrow}  \\
	 -\tilde{f}_{i,\downarrow}  \\
	 \phantom{-}  \tilde{f}_{i,\uparrow}  
\end{pmatrix}
\end{equation}
with $T^2 = -1$.   Kramers  theorem thus guaranties that the eigenvalues of the fermion determinant matrix   come in complex conjugate pairs \cite{Wu04}  such that the fermion weights are always positive. 

\subsection{Histograms}

\begin{figure}
\centering
\centerline{\includegraphics[width=0.4\textwidth]{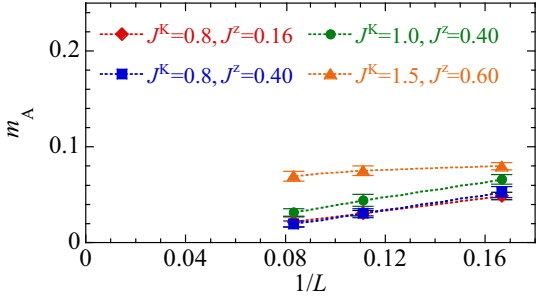}}
\caption{\label{fig:Sup-Histo-0}
Magnetization $m_{A}$ of the local moments along the $z$ axis belonging to the A sublattice ( $m_{A}=m_{B}$) for the z-PKS phase.
Here, $T=0.04$.
}
\end{figure}
\begin{figure}
\centering
\centerline{\includegraphics[width=0.325\textwidth]{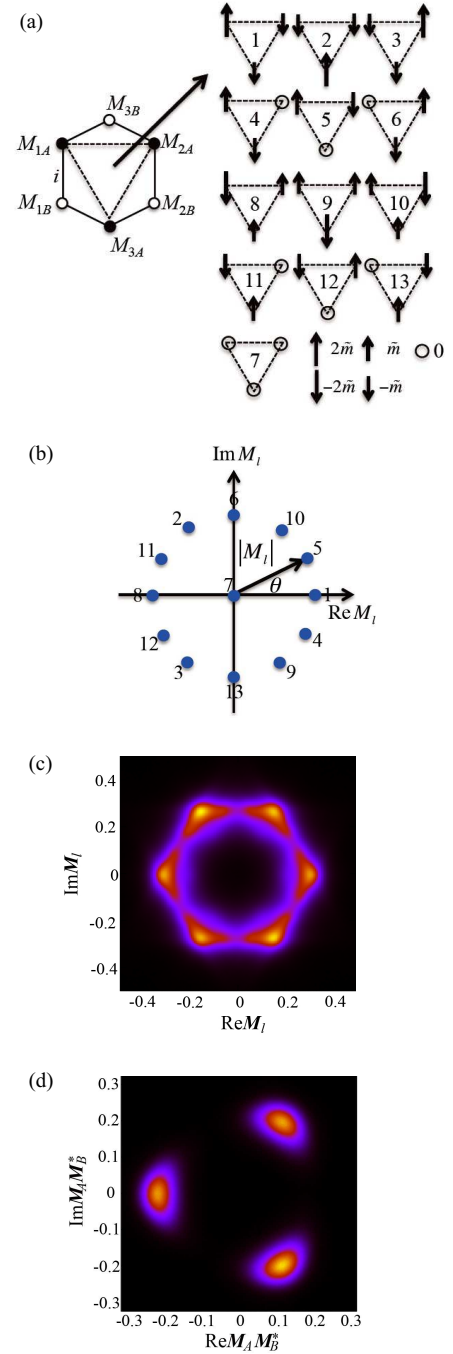}}
\caption{\label{fig:Sup-Histo-1}
(a) Schematic spin structure of local moments on the triangle belonging to A sublattice for the z-PKS phase and 
(b) schematic probability distribution of $\bm M_l$ corresponding to the thirteen patterns.
Typical QMC histogram results of (c) $\bm M_l$ and (d) ${\bm M}_A {\bm M}^{*}_B$ for the z-PKS phase.}
\end{figure}

In this section we will detail the QMC histogram calculation presented  in the paper  and show how to extract the spin structure of the  z-PKS phase.
We first introduce the complex number 
\begin{eqnarray}
{\bm M}_{l}=M_{1l}e^{i0}+M_{2l}e^{i\frac{2\pi}{3}}+M_{3l}e^{i\frac{4\pi}{3}},
\label{eq:M}
\end{eqnarray}
where $M_{il}$  is  the  magnetization  along the $z$ axis at sites $i$ of the triangle  defining sublattice $l (=A, B) $ in the  six-sites hexagon forming the unit cell of the z-PKS order. 
As shown in Fig.~\ref{fig:Sup-Histo-0}, the magnetization on each sublattice $m_{l}= \sqrt{\frac{1}{N^2}\sum_{i,j}\langle M_{il} M_{jl}\rangle}$ obtained from finite-size scaling is found to be small,  and  even almost vanishing in  the z-PKS phase. Note that a fully polarized state has $m_{l} = 0.5$.  We will henceforth  assume that $m_{l}  \simeq 0 $.  In the z-PKS phase,  U(1) symmetry is not broken, and Kondo screening will allow for a variable  site dependent magnitude of the local moment. Inspiring ourselves from the Ising model in a transverse field on the triangular lattice \cite{Moessner01b} and with the aforementioned constraint   $m_{l}  \simeq 0 $ we will consider the following thirteen patterns (see Fig.~\ref{fig:Sup-Histo-1}(a)).  $\bm M_l$ takes a distinct value for each of the  thirteen patterns (see Fig.~\ref{fig:Sup-Histo-1}(b)).
In particular and with  $\bm M_{l}\equiv |{\bm M_{l}}|e^{i\theta}$   the nonmagnetic state  $(M_{1l},M_{2l},M_{3l})=(0,0,0)$  corresponds to  $|{\bm M_{l}}|=0$,   the six-fold degenerate magnetic state  $(M_{1l},M_{2l},M_{3l})=\tilde{m}(2,-1,-1)$ to $\theta=\frac{2n\pi}{6}$  and $|{\bm M_{l}}|\neq0$, and the  six-fold degenerate magnetic state $(M_{1l},M_{2l},M_{3l})=\tilde{m}(1,-1,0)$
to $\theta=\frac{2(n+1)\pi}{6}$ and $|{\bm M_{l}}|\neq0$.  Here  $n=0,2,3,...,6$ and $\tilde{m}$ is a constant.
A typical result for the QMC histogram of  $\bm M_l$ in the z-PKS phase is presented in Fig.~\ref{fig:Sup-Histo-1}(c).
The result shows six peaks at $\bm M_l\sim \pm 0.3+0i $, $0.15 \pm 0.3i$, and $-0.15 \pm 0.3i$ which correspond to the
the six-fold degenerate magnetic state of $(M_{1l},M_{2l},M_{3l})=\tilde{m}(2,-1,-1)$.
Indeed, the magnetization estimated from these peaks is consistent with the six-fold degenerate states 
\begin{eqnarray}
(M_{1l},M_{2l},M_{3l})\sim(+0.2,-0.1,-0.1).
\label{eq:M-result}
\end{eqnarray}

To  investigate correlations between the two sublattices we   consider
\begin{eqnarray}
{\bm M}_A {\bm M}^{*}_B & = & \left( M_{1A}e^{i0}+M_{2A}e^{i\frac{2\pi}{3}}+M_{3A}e^{i\frac{4\pi}{3}}\right) \\
&\times&\left( M_{1B}e^{-i0}+M_{2B}e^{i\frac{-2\pi}{3}}+M_{3B}e^{i\frac{-4\pi}{3}}\right) \nonumber
\label{eq:MM}
\end{eqnarray}
A typical result for the QMC histogram of ${\bm M}_A {\bm M}^{*}_B$ in the z-PKS phase is shown in Fig.~\ref{fig:Sup-Histo-1}(d).
The result shows three peaks at ${\bm M}_A {\bm M}^{*}_B\sim -0.2+0i $ and $0.1 \pm 0.2i$. Each peak has a two fold degeneracy  since ${\bm M}_A {\bm M}^{*}_B$  is invariant under ${\bm M}_A \rightarrow -{\bm M}_A$ and 
${\bm M}_B\rightarrow -{\bm M}_B$.  Thereby the ground state has the same  6-fold degeneracy as  on a single sublattice such that magnetic ordering between  sublattices is locked in. 
For example, consider the previously observed pattern  $(M_{1A},M_{2A},M_{3A})\sim(+0.2,-0.1,-0.1)$  then 
$(M_{1B},M_{2B},M_{3B})\sim(-0.2,+0.1,+0.1)$     will give a signal approximately  at $-0.2+0i $ in the  histogram of  ${\bm M}_A {\bm M}^{*}_B $.   C$_3$    rotations of this structure will account for  the two other peaks in the  histogram. 

\subsection{Mass Terms}

\begin{figure}
\centering
\centerline{\includegraphics[width=0.3\textwidth]{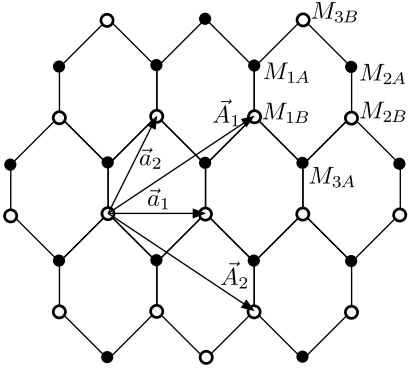}}
\caption{\label{fig:Honeycomb}
The honeycomb lattice with enhanced unit cell  spanned by the lattice vectors $\pmb{A}_1$ and $\pmb{A}_2$   harboring the z-PKS phase.}
\end{figure}

To investigate  how the z-PKS magnetic ordering  induces a mass in the Dirac fermions we  consider  the model Hamiltonian $\hat{H} = \hat{H}_{\text{t}}  + 
\hat{H}_{zPKS} $  with 
\begin{eqnarray}
	\hat{H}_{\text{t}}   & =  &- t \sum_{\pmb{r}} \hat{b}^{\dagger}_{\pmb{r}} \left(  \hat{a}^{\phantom\dagger}_{\pmb{r}}   + 
	\hat{a}^{\phantom\dagger}_{\pmb{r} - \pmb{a}_2 }    + \hat{a}^{\phantom\dagger}_{\pmb{r}  + \pmb{a}_1  - \pmb{a}_2 }  \right)  + h.c. 
       \text{ and } \nonumber \\ 
	\hat{H}_{\text{z-PKS}} & = &   \frac{J^{K}}{2} \sum_{\pmb{R}}  \left(  
	    M_{1A} \hat{a}^{\dagger}_{\pmb{R} }\sigma_{z}  \hat{a}^{\phantom\dagger}_{\pmb{R}}  + M_{2A} \hat{a}^{\dagger}_{\pmb{R} +\pmb{a}_1}\sigma_{z}  \hat{a}^{\phantom\dagger}_{\pmb{R}+\pmb{a}_1} \right.  \nonumber \\
	    &  &   \left.  M_{3A}   \hat{a}^{\dagger}_{\pmb{R} +\pmb{a}_2}\sigma_{z}  \hat{a}^{\phantom\dagger}_{\pmb{R}+\pmb{a}_2}   \right)     + a \leftrightarrow b.
\end{eqnarray}
In the above  (see Fig.~\ref{fig:Honeycomb}), $\pmb{r}  = n_1 \pmb{a}_1 + n_2 \pmb{a}_2 $,  $  \pmb{R}  = \tilde{n}_1 \pmb{A}_1  + \tilde{n}_2 \pmb{A}_2 $, $M_{1A}$  and equivalent forms corresponds to the magnetization of the local moments in the z-PKS phase  and finally $ \hat{a}^{\dagger}_{\pmb{r}}$ and $\hat{b}^{\dagger}_{\pmb{r}}$ are two components spinors encoding the  spin degree of freedom.     Fourier transformation, 
$\hat{a}^{\dagger}_{\pmb{k}} = \frac{1}{\sqrt{N}} \sum_{r} e^{i \pmb{k} \cdot \pmb{r}}   \hat{a}^{\dagger}_{\pmb{r}} $   with $N$ the number of unit cells of the  Honeycomb lattice (spanned by the lattice vectors $\pmb{a}_1$, $ \pmb{a}_2$) and expansion around  the two Dirac points,  
\begin{equation}
	\pmb{K}   = \pm \left( \frac{4}{3} \pmb{b}_1  +  \frac{2}{3} \pmb{b}_2  \right)   \text{ with }      \pmb{a}_i \cdot \pmb{b}_j  = 2 \pi \delta_{i,j}
\end{equation}
defines the low energy modes:
\begin{equation}
	\hat{c}^{\dagger}_{s=\left\{ A,B \right\},v=\left\{ \pmb{K}, -\pmb{K} \right\}, \sigma=\left\{ \uparrow,\downarrow \right\} } (\boldsymbol{p}). 
\end{equation}
Here $s$ denotes the sub-lattice index, $v$ the valley index, and $\sigma$ the physical spin. The corresponding Pauli matrices will be denoted by 
$\boldsymbol{\tau}$, $\boldsymbol{ \mu}$ and $\boldsymbol{\sigma}$ respectively.   Finally, $\boldsymbol{p}$ is the momentum measured with respect to  the valley momentum.  The canonical transformation
\begin{equation}
	\hat{c}^{\dagger}  =  \hat{\Psi}^{\dagger} \left(  \tau^{y} P_{+}  +  P_{-} \right)  \text{  with }   P_{\pm} = \frac{1}{2} \left( \mu^0 \pm \mu^z \right) 
\end{equation} 
yields the Dirac Hamiltonian 
\begin{equation}
  \hat{H} _{\text{Dirac}}   =   \sum_{\pmb{p}}  \hat{\Psi}^{\dagger}(\pmb{p})      \left[  p_x \tau^{x}      + p_y \tau^{y} \right] 
 \hat{\Psi}(\pmb{p})    
\end{equation}
where we have set the velocity $ v_F= \sqrt{3} a t/2 $  to unity.  Note that in this form, the Dirac Hamiltonian in (2+1)D   has a manifest O(8) symmetry. 

We now turn our attention to $\hat{H}_{\text{z-PKS}} $.  Using 
\begin{equation}
  \frac{3}{N}  \sum_{\pmb{R}}     e^{i\pmb{k} \cdot \pmb{R}} =  \delta_{\pmb{k},\pmb{0}}  + \delta_{\pmb{k},2\pmb{K}}  + \delta_{\pmb{k},-2\pmb{K}} 
\end{equation} 
we obtain:
\begin{eqnarray}
\label{Eq:Hpks}
	\hat{H}_{\text{z-PKS}} =& &  \frac{J^{\text{K}}}{6} \left( \sum_{n=1}^{3} M_{n A}\right)  \sum_{\pmb{k}} \hat{a}^{\dagger}_{\pmb{k} }\sigma_{z}  \hat{a}^{\phantom\dagger}_{\pmb{k}}  + a \leftrightarrow b     \\
	& + &   \frac{J^{\text{K}}}{6} \sum_{\pmb{k}}  \left(   M_A \hat{a}^{\dagger}_{\pmb{k} }\sigma_{z}  \hat{a}^{\phantom\dagger}_{\pmb{k} + 2 \pmb{K}}   + h.c. \right)    +  a \leftrightarrow b  \nonumber
\end{eqnarray}
Here, 
\begin{equation}
	 M_A =  M_{1A} + M_{2A}e^{i 4 \pi/3} + M_{3A} e^{i 2 \pi/3}
\end{equation}

Assume  that $ \left( \sum_{n=1}^{3} M_{nA}\right)   = m \neq 0 $ and that the total magnetization per  hexagon vanishes, (i.e. $\left( \sum_{n=1}^{3} M_{n,B}\right)   = - m$ ),  then  first term of Eq.~(\ref{Eq:Hpks})  corresponds to a mass term. In particular, expanding around the Dirac points,   gives:
\begin{equation}
\hat{H}_{\text{z-PKS}}   \simeq  \frac{J^{\text{K}}m}{6} \sum_{\pmb{p}}  \hat{\Psi}^{\dagger}(\pmb{p})\sigma^{z} \tau^{z} \mu^{z}   \hat{\Psi}(\pmb{p}) 
\end{equation}
which corresponds to a mass term. 

In the   parameter range where we have  carried out our QMC  simulations  we often found that   $ \left( \sum_{n=1}^{3} M_{nA}\right)   \simeq 0  $ and that the spin ordering observed in the  z-PKS phase  is  consistent  with 
$ (M_{1A}, M_{2A},M_{3A})  = \tilde{m} (2,-1,-1) $ and $ (M_{1B}, M_{2B},M_{3B}) = \tilde{m}(-2,1,1) $. In this case $M_A  = 1$, and  since the first term of Eq.~(\ref{Eq:Hpks}) vanishes,
\begin{equation} 
\hat{H}_{\text{z-PKS}}   \simeq  \frac{J^{\text{K}} \tilde{m}}{6} \sum_{\pmb{p}}  \hat{\Psi}^{\dagger}(\pmb{p})  \sigma^{z}   \tau^{x}   \hat{\Psi}(\pmb{p}).
\end{equation}
Solving for the spectrum of $ \hat{H}_{\text{Dirac}} +  \hat{H}_{\text{z-PKS}}$   gives
\begin{equation}
	E(\pmb{p})  =  \pm \sqrt{  \pmb{p}^2 + \left( \frac{J^{\text{K}} \tilde{m}}{6}  \right)^2  \pm 2 \left|  p_x  \frac{J^{\text{K}}\tilde{m}}{6}   \right| }
\end{equation}
  with zero modes at 
  \begin{equation}
  	 \pmb{p}  = \left( \pm  \frac{J^{\text{K}}\tilde{m}}{6} , 0 \right)
  \end{equation}
Thereby,  the magnetic ordering of the z-PKS phase does not necessarily open a charge gap in the Dirac fermions. 
In triggers a  nematic transition where the Dirac points meander away from the origin. 


\begin{thebibliography}{40}%
\makeatletter
\providecommand \@ifxundefined [1]{%
 \@ifx{#1\undefined}
}%
\providecommand \@ifnum [1]{%
 \ifnum #1\expandafter \@firstoftwo
 \else \expandafter \@secondoftwo
 \fi
}%
\providecommand \@ifx [1]{%
 \ifx #1\expandafter \@firstoftwo
 \else \expandafter \@secondoftwo
 \fi
}%
\providecommand \natexlab [1]{#1}%
\providecommand \enquote  [1]{``#1''}%
\providecommand \bibnamefont  [1]{#1}%
\providecommand \bibfnamefont [1]{#1}%
\providecommand \citenamefont [1]{#1}%
\providecommand \href@noop [0]{\@secondoftwo}%
\providecommand \href [0]{\begingroup \@sanitize@url \@href}%
\providecommand \@href[1]{\@@startlink{#1}\@@href}%
\providecommand \@@href[1]{\endgroup#1\@@endlink}%
\providecommand \@sanitize@url [0]{\catcode `\\12\catcode `\$12\catcode
  `\&12\catcode `\#12\catcode `\^12\catcode `\_12\catcode `\%12\relax}%
\providecommand \@@startlink[1]{}%
\providecommand \@@endlink[0]{}%
\providecommand \url  [0]{\begingroup\@sanitize@url \@url }%
\providecommand \@url [1]{\endgroup\@href {#1}{\urlprefix }}%
\providecommand \urlprefix  [0]{URL }%
\providecommand \Eprint [0]{\href }%
\providecommand \doibase [0]{http://dx.doi.org/}%
\providecommand \selectlanguage [0]{\@gobble}%
\providecommand \bibinfo  [0]{\@secondoftwo}%
\providecommand \bibfield  [0]{\@secondoftwo}%
\providecommand \translation [1]{[#1]}%
\providecommand \BibitemOpen [0]{}%
\providecommand \bibitemStop [0]{}%
\providecommand \bibitemNoStop [0]{.\EOS\space}%
\providecommand \EOS [0]{\spacefactor3000\relax}%
\providecommand \BibitemShut  [1]{\csname bibitem#1\endcsname}%
\let\auto@bib@innerbib\@empty
\bibitem [{\citenamefont {Kitaev}(2003)}]{Kitaev03}%
  \BibitemOpen
  \bibfield  {author} {\bibinfo {author} {\bibfnamefont {A.}~\bibnamefont
  {Kitaev}},\ }\href {\doibase http://dx.doi.org/10.1016/S0003-4916(02)00018-0}
  {\bibfield  {journal} {\bibinfo  {journal} {Annals of Physics}\ }\textbf
  {\bibinfo {volume} {303}},\ \bibinfo {pages} {2 } (\bibinfo {year}
  {2003})}\BibitemShut {NoStop}%
\bibitem [{\citenamefont {Hubbard}(1963)}]{Hubbard63}%
  \BibitemOpen
  \bibfield  {author} {\bibinfo {author} {\bibfnamefont {J.}~\bibnamefont
  {Hubbard}},\ }\href {\doibase 10.1098/rspa.1963.0204} {\bibfield  {journal}
  {\bibinfo  {journal} {Proceedings of the Royal Society of London A:
  Mathematical, Physical and Engineering Sciences}\ }\textbf {\bibinfo {volume}
  {276}},\ \bibinfo {pages} {238} (\bibinfo {year} {1963})}\BibitemShut
  {NoStop}%
\bibitem [{\citenamefont {Henelius}\ and\ \citenamefont
  {Sandvik}(2000)}]{Sandvik99b1}%
  \BibitemOpen
  \bibfield  {author} {\bibinfo {author} {\bibfnamefont {P.}~\bibnamefont
  {Henelius}}\ and\ \bibinfo {author} {\bibfnamefont {A.}~\bibnamefont
  {Sandvik}},\ }\href@noop {} {\bibfield  {journal} {\bibinfo  {journal} {Phys.
  Rev. B}\ }\textbf {\bibinfo {volume} {62}},\ \bibinfo {pages} {1102}
  (\bibinfo {year} {2000})}\BibitemShut {NoStop}%
\bibitem [{\citenamefont {Loh}\ \emph {et~al.}(1990)\citenamefont {Loh},
  \citenamefont {Gubernatis}, \citenamefont {Scalettar}, \citenamefont {White},
  \citenamefont {Scalapino},\ and\ \citenamefont {Sugar}}]{Loh1990}%
  \BibitemOpen
  \bibfield  {author} {\bibinfo {author} {\bibfnamefont {E.~Y.}\ \bibnamefont
  {Loh}}, \bibinfo {author} {\bibfnamefont {J.~E.}\ \bibnamefont {Gubernatis}},
  \bibinfo {author} {\bibfnamefont {R.~T.}\ \bibnamefont {Scalettar}}, \bibinfo
  {author} {\bibfnamefont {S.~R.}\ \bibnamefont {White}}, \bibinfo {author}
  {\bibfnamefont {D.~J.}\ \bibnamefont {Scalapino}}, \ and\ \bibinfo {author}
  {\bibfnamefont {R.~L.}\ \bibnamefont {Sugar}},\ }\href {\doibase
  10.1103/PhysRevB.41.9301} {\bibfield  {journal} {\bibinfo  {journal} {Phys.
  Rev. B}\ }\textbf {\bibinfo {volume} {41}},\ \bibinfo {pages} {9301}
  (\bibinfo {year} {1990})}\BibitemShut {NoStop}%
\bibitem [{\citenamefont {Troyer}\ and\ \citenamefont
  {Wiese}(2005)}]{Troyer05}%
  \BibitemOpen
  \bibfield  {author} {\bibinfo {author} {\bibfnamefont {M.}~\bibnamefont
  {Troyer}}\ and\ \bibinfo {author} {\bibfnamefont {U.-J.}\ \bibnamefont
  {Wiese}},\ }\href {\doibase 10.1103/PhysRevLett.94.170201} {\bibfield
  {journal} {\bibinfo  {journal} {Phys. Rev. Lett.}\ }\textbf {\bibinfo
  {volume} {94}},\ \bibinfo {pages} {170201} (\bibinfo {year}
  {2005})}\BibitemShut {NoStop}%
\bibitem [{\citenamefont {Blankenbecler}\ \emph {et~al.}(1981)\citenamefont
  {Blankenbecler}, \citenamefont {Scalapino},\ and\ \citenamefont
  {Sugar}}]{Blankenbecler81}%
  \BibitemOpen
  \bibfield  {author} {\bibinfo {author} {\bibfnamefont {R.}~\bibnamefont
  {Blankenbecler}}, \bibinfo {author} {\bibfnamefont {D.~J.}\ \bibnamefont
  {Scalapino}}, \ and\ \bibinfo {author} {\bibfnamefont {R.~L.}\ \bibnamefont
  {Sugar}},\ }\href {\doibase 10.1103/PhysRevD.24.2278} {\bibfield  {journal}
  {\bibinfo  {journal} {Phys. Rev. D}\ }\textbf {\bibinfo {volume} {24}},\
  \bibinfo {pages} {2278} (\bibinfo {year} {1981})}\BibitemShut {NoStop}%
\bibitem [{\citenamefont {White}\ \emph {et~al.}(1989)\citenamefont {White},
  \citenamefont {Scalapino}, \citenamefont {Sugar}, \citenamefont {Loh},
  \citenamefont {Gubernatis},\ and\ \citenamefont {Scalettar}}]{White89}%
  \BibitemOpen
  \bibfield  {author} {\bibinfo {author} {\bibfnamefont {S.}~\bibnamefont
  {White}}, \bibinfo {author} {\bibfnamefont {D.}~\bibnamefont {Scalapino}},
  \bibinfo {author} {\bibfnamefont {R.}~\bibnamefont {Sugar}}, \bibinfo
  {author} {\bibfnamefont {E.}~\bibnamefont {Loh}}, \bibinfo {author}
  {\bibfnamefont {J.}~\bibnamefont {Gubernatis}}, \ and\ \bibinfo {author}
  {\bibfnamefont {R.}~\bibnamefont {Scalettar}},\ }\href {\doibase
  10.1103/PhysRevB.40.506} {\bibfield  {journal} {\bibinfo  {journal} {Phys.
  Rev. B}\ }\textbf {\bibinfo {volume} {40}},\ \bibinfo {pages} {506} (\bibinfo
  {year} {1989})}\BibitemShut {NoStop}%
\bibitem [{\citenamefont {Assaad}\ and\ \citenamefont
  {Evertz}(2008)}]{Assaad08_rev}%
  \BibitemOpen
  \bibfield  {author} {\bibinfo {author} {\bibfnamefont {F.}~\bibnamefont
  {Assaad}}\ and\ \bibinfo {author} {\bibfnamefont {H.}~\bibnamefont
  {Evertz}},\ }in\ \href {\doibase 10.1007/978-3-540-74686-7_10} {\emph
  {\bibinfo {booktitle} {Computational Many-Particle Physics}}},\ \bibinfo
  {series} {Lecture Notes in Physics}, Vol.\ \bibinfo {volume} {739},\ \bibinfo
  {editor} {edited by\ \bibinfo {editor} {\bibfnamefont {H.}~\bibnamefont
  {Fehske}}, \bibinfo {editor} {\bibfnamefont {R.}~\bibnamefont {Schneider}}, \
  and\ \bibinfo {editor} {\bibfnamefont {A.}~\bibnamefont {Wei{\ss}e}}}\
  (\bibinfo  {publisher} {Springer},\ \bibinfo {address} {Berlin Heidelberg},\
  \bibinfo {year} {2008})\ pp.\ \bibinfo {pages} {277--356}\BibitemShut
  {NoStop}%
\bibitem [{\citenamefont {Moessner}\ and\ \citenamefont
  {Sondhi}(2001{\natexlab{a}})}]{Moessner01}%
  \BibitemOpen
  \bibfield  {author} {\bibinfo {author} {\bibfnamefont {R.}~\bibnamefont
  {Moessner}}\ and\ \bibinfo {author} {\bibfnamefont {S.~L.}\ \bibnamefont
  {Sondhi}},\ }\href {\doibase 10.1103/PhysRevLett.86.1881} {\bibfield
  {journal} {\bibinfo  {journal} {Phys. Rev. Lett.}\ }\textbf {\bibinfo
  {volume} {86}},\ \bibinfo {pages} {1881} (\bibinfo {year}
  {2001}{\natexlab{a}})}\BibitemShut {NoStop}%
\bibitem [{\citenamefont {Isakov}\ \emph {et~al.}(2011)\citenamefont {Isakov},
  \citenamefont {Hastings},\ and\ \citenamefont {Melko}}]{Isakov11}%
  \BibitemOpen
  \bibfield  {author} {\bibinfo {author} {\bibfnamefont {S.~V.}\ \bibnamefont
  {Isakov}}, \bibinfo {author} {\bibfnamefont {M.~B.}\ \bibnamefont
  {Hastings}}, \ and\ \bibinfo {author} {\bibfnamefont {R.~G.}\ \bibnamefont
  {Melko}},\ }\href {\doibase 10.1038/nphys2036} {\bibfield  {journal}
  {\bibinfo  {journal} {Nature Phys.}\ }\textbf {\bibinfo {volume} {7}},\
  \bibinfo {pages} {772} (\bibinfo {year} {2011})}\BibitemShut {NoStop}%
\bibitem [{\citenamefont {Aulbach}\ \emph {et~al.}(2015)\citenamefont
  {Aulbach}, \citenamefont {Assaad},\ and\ \citenamefont
  {Potthoff}}]{Aulbach2015}%
  \BibitemOpen
  \bibfield  {author} {\bibinfo {author} {\bibfnamefont {M.~W.}\ \bibnamefont
  {Aulbach}}, \bibinfo {author} {\bibfnamefont {F.~F.}\ \bibnamefont {Assaad}},
  \ and\ \bibinfo {author} {\bibfnamefont {M.}~\bibnamefont {Potthoff}},\
  }\href {\doibase 10.1103/PhysRevB.92.235131} {\bibfield  {journal} {\bibinfo
  {journal} {Phys. Rev. B}\ }\textbf {\bibinfo {volume} {92}},\ \bibinfo
  {pages} {235131} (\bibinfo {year} {2015})}\BibitemShut {NoStop}%
\bibitem [{\citenamefont {{Pixley}}\ \emph {et~al.}(2015)\citenamefont
  {{Pixley}}, \citenamefont {{Yu}}, \citenamefont {{Paschen}},\ and\
  \citenamefont {{Si}}}]{Pixley2015}%
  \BibitemOpen
  \bibfield  {author} {\bibinfo {author} {\bibfnamefont {J.~H.}\ \bibnamefont
  {{Pixley}}}, \bibinfo {author} {\bibfnamefont {R.}~\bibnamefont {{Yu}}},
  \bibinfo {author} {\bibfnamefont {S.}~\bibnamefont {{Paschen}}}, \ and\
  \bibinfo {author} {\bibfnamefont {Q.}~\bibnamefont {{Si}}},\ }\href@noop {}
  {\bibfield  {journal} {\bibinfo  {journal} {ArXiv e-prints}\ } (\bibinfo
  {year} {2015})},\ \Eprint {http://arxiv.org/abs/1509.02907}
  {arXiv:1509.02907} \BibitemShut {NoStop}%
\bibitem [{\citenamefont {{Pixley}}\ \emph {et~al.}(2016)\citenamefont
  {{Pixley}}, \citenamefont {{Lee}}, \citenamefont {{Brandom}},\ and\
  \citenamefont {{Parameswaran}}}]{Pixley2016}%
  \BibitemOpen
  \bibfield  {author} {\bibinfo {author} {\bibfnamefont {J.~H.}\ \bibnamefont
  {{Pixley}}}, \bibinfo {author} {\bibfnamefont {S.}~\bibnamefont {{Lee}}},
  \bibinfo {author} {\bibfnamefont {B.}~\bibnamefont {{Brandom}}}, \ and\
  \bibinfo {author} {\bibfnamefont {S.~A.}\ \bibnamefont {{Parameswaran}}},\
  }\href@noop {} {\bibfield  {journal} {\bibinfo  {journal} {ArXiv e-prints}\ }
  (\bibinfo {year} {2016})},\ \Eprint {http://arxiv.org/abs/1609.04023}
  {arXiv:1609.04023} \BibitemShut {NoStop}%
\bibitem [{\citenamefont {Coleman}\ and\ \citenamefont
  {Nevidomskyy}(2010)}]{Coleman10}%
  \BibitemOpen
  \bibfield  {author} {\bibinfo {author} {\bibfnamefont {P.}~\bibnamefont
  {Coleman}}\ and\ \bibinfo {author} {\bibfnamefont {A.~H.}\ \bibnamefont
  {Nevidomskyy}},\ }\href@noop {} {\bibfield  {journal} {\bibinfo  {journal}
  {J. Low Temp. Phys.}\ }\textbf {\bibinfo {volume} {161}},\ \bibinfo {pages}
  {182} (\bibinfo {year} {2010})}\BibitemShut {NoStop}%
\bibitem [{\citenamefont {Motome}\ \emph {et~al.}(2010)\citenamefont {Motome},
  \citenamefont {Nakamikawa}, \citenamefont {Yamaji},\ and\ \citenamefont
  {Udagawa}}]{Motome2010}%
  \BibitemOpen
  \bibfield  {author} {\bibinfo {author} {\bibfnamefont {Y.}~\bibnamefont
  {Motome}}, \bibinfo {author} {\bibfnamefont {K.}~\bibnamefont {Nakamikawa}},
  \bibinfo {author} {\bibfnamefont {Y.}~\bibnamefont {Yamaji}}, \ and\ \bibinfo
  {author} {\bibfnamefont {M.}~\bibnamefont {Udagawa}},\ }\href {\doibase
  10.1103/PhysRevLett.105.036403} {\bibfield  {journal} {\bibinfo  {journal}
  {Phys. Rev. Lett.}\ }\textbf {\bibinfo {volume} {105}},\ \bibinfo {pages}
  {036403} (\bibinfo {year} {2010})}\BibitemShut {NoStop}%
\bibitem [{\citenamefont {Ishizuka}\ and\ \citenamefont
  {Motome}(2013)}]{Ishizuka2013}%
  \BibitemOpen
  \bibfield  {author} {\bibinfo {author} {\bibfnamefont {H.}~\bibnamefont
  {Ishizuka}}\ and\ \bibinfo {author} {\bibfnamefont {Y.}~\bibnamefont
  {Motome}},\ }\href {\doibase 10.1103/PhysRevB.88.081105} {\bibfield
  {journal} {\bibinfo  {journal} {Phys. Rev. B}\ }\textbf {\bibinfo {volume}
  {88}},\ \bibinfo {pages} {081105} (\bibinfo {year} {2013})}\BibitemShut
  {NoStop}%
\bibitem [{\citenamefont {Schattner}\ \emph {et~al.}(2016)\citenamefont
  {Schattner}, \citenamefont {Lederer}, \citenamefont {Kivelson},\ and\
  \citenamefont {Berg}}]{Schattner15}%
  \BibitemOpen
  \bibfield  {author} {\bibinfo {author} {\bibfnamefont {Y.}~\bibnamefont
  {Schattner}}, \bibinfo {author} {\bibfnamefont {S.}~\bibnamefont {Lederer}},
  \bibinfo {author} {\bibfnamefont {S.~A.}\ \bibnamefont {Kivelson}}, \ and\
  \bibinfo {author} {\bibfnamefont {E.}~\bibnamefont {Berg}},\ }\href {\doibase
  10.1103/PhysRevX.6.031028} {\bibfield  {journal} {\bibinfo  {journal} {Phys.
  Rev. X}\ }\textbf {\bibinfo {volume} {6}},\ \bibinfo {pages} {031028}
  (\bibinfo {year} {2016})}\BibitemShut {NoStop}%
\bibitem [{\citenamefont {{Liu}}\ \emph {et~al.}()\citenamefont {{Liu}},
  \citenamefont {{Xu}}, \citenamefont {{Qi}}, \citenamefont {{Sun}},\ and\
  \citenamefont {{Meng}}}]{liu_frust17}%
  \BibitemOpen
  \bibfield  {author} {\bibinfo {author} {\bibfnamefont {Z.~H.}\ \bibnamefont
  {{Liu}}}, \bibinfo {author} {\bibfnamefont {X.~Y.}\ \bibnamefont {{Xu}}},
  \bibinfo {author} {\bibfnamefont {Y.}~\bibnamefont {{Qi}}}, \bibinfo {author}
  {\bibfnamefont {K.}~\bibnamefont {{Sun}}}, \ and\ \bibinfo {author}
  {\bibfnamefont {Z.~Y.}\ \bibnamefont {{Meng}}},\ }\href@noop {} {\bibfield
  {journal} {\bibinfo  {journal} {ArXiv e-prints}\ }}\Eprint
  {http://arxiv.org/abs/1706.10004} {arXiv:1706.10004} \BibitemShut {NoStop}%
\bibitem [{\citenamefont {Assaad}\ and\ \citenamefont
  {Grover}(2016)}]{grover_assaad16}%
  \BibitemOpen
  \bibfield  {author} {\bibinfo {author} {\bibfnamefont {F.~F.}\ \bibnamefont
  {Assaad}}\ and\ \bibinfo {author} {\bibfnamefont {T.}~\bibnamefont
  {Grover}},\ }\href {\doibase 10.1103/PhysRevX.6.041049} {\bibfield  {journal}
  {\bibinfo  {journal} {Phys. Rev. X}\ }\textbf {\bibinfo {volume} {6}},\
  \bibinfo {pages} {041049} (\bibinfo {year} {2016})}\BibitemShut {NoStop}%
\bibitem [{\citenamefont {Gazit}\ \emph {et~al.}(2017)\citenamefont {Gazit},
  \citenamefont {Randeria},\ and\ \citenamefont {Vishwanath}}]{gazit2016}%
  \BibitemOpen
  \bibfield  {author} {\bibinfo {author} {\bibfnamefont {S.}~\bibnamefont
  {Gazit}}, \bibinfo {author} {\bibfnamefont {M.}~\bibnamefont {Randeria}}, \
  and\ \bibinfo {author} {\bibfnamefont {A.}~\bibnamefont {Vishwanath}},\
  }\href {http://dx.doi.org/10.1038/nphys4028} {\bibfield  {journal} {\bibinfo
  {journal} {Nat Phys}\ }\textbf {\bibinfo {volume} {13}},\ \bibinfo {pages}
  {484} (\bibinfo {year} {2017})}\BibitemShut {NoStop}%
\bibitem [{\citenamefont {Abrikosov}(1965)}]{Abrikosov1965electron}%
  \BibitemOpen
  \bibfield  {author} {\bibinfo {author} {\bibfnamefont {A.}~\bibnamefont
  {Abrikosov}},\ }\href@noop {} {\bibfield  {journal} {\bibinfo  {journal}
  {Physics}\ }\textbf {\bibinfo {volume} {2}},\ \bibinfo {pages} {5} (\bibinfo
  {year} {1965})}\BibitemShut {NoStop}%
\bibitem [{\citenamefont {Wu}\ and\ \citenamefont {Zhang}(2005)}]{Wu04}%
  \BibitemOpen
  \bibfield  {author} {\bibinfo {author} {\bibfnamefont {C.}~\bibnamefont
  {Wu}}\ and\ \bibinfo {author} {\bibfnamefont {S.-C.}\ \bibnamefont {Zhang}},\
  }\href {\doibase 10.1103/PhysRevB.71.155115} {\bibfield  {journal} {\bibinfo
  {journal} {Phys. Rev. B}\ }\textbf {\bibinfo {volume} {71}},\ \bibinfo
  {pages} {155115} (\bibinfo {year} {2005})}\BibitemShut {NoStop}%
\bibitem [{\citenamefont {Assaad}(1999)}]{Assaad99a}%
  \BibitemOpen
  \bibfield  {author} {\bibinfo {author} {\bibfnamefont {F.~F.}\ \bibnamefont
  {Assaad}},\ }\href@noop {} {\bibfield  {journal} {\bibinfo  {journal} {Phys.
  Rev. Lett.}\ }\textbf {\bibinfo {volume} {83}},\ \bibinfo {pages} {796}
  (\bibinfo {year} {1999})}\BibitemShut {NoStop}%
\bibitem [{\citenamefont {Capponi}\ and\ \citenamefont
  {Assaad}(2001)}]{Capponi00}%
  \BibitemOpen
  \bibfield  {author} {\bibinfo {author} {\bibfnamefont {S.}~\bibnamefont
  {Capponi}}\ and\ \bibinfo {author} {\bibfnamefont {F.~F.}\ \bibnamefont
  {Assaad}},\ }\href@noop {} {\bibfield  {journal} {\bibinfo  {journal} {Phys.
  Rev. B}\ }\textbf {\bibinfo {volume} {63}},\ \bibinfo {pages} {155114}
  (\bibinfo {year} {2001})}\BibitemShut {NoStop}%
\bibitem [{\citenamefont {Doniach}(1977)}]{Doniach77}%
  \BibitemOpen
  \bibfield  {author} {\bibinfo {author} {\bibfnamefont {S.}~\bibnamefont
  {Doniach}},\ }\href@noop {} {\bibfield  {journal} {\bibinfo  {journal}
  {Physica B}\ }\textbf {\bibinfo {volume} {91}},\ \bibinfo {pages} {231}
  (\bibinfo {year} {1977})}\BibitemShut {NoStop}%
\bibitem [{\citenamefont {Hewson}(1997)}]{Hewson}%
  \BibitemOpen
  \bibfield  {author} {\bibinfo {author} {\bibfnamefont {A.~C.}\ \bibnamefont
  {Hewson}},\ }\href@noop {} {\emph {\bibinfo {title} {The Kondo Problem to
  Heavy Fermions}}},\ Cambridge Studies in Magnetism\ (\bibinfo  {publisher}
  {Cambridge Universiy Press},\ \bibinfo {address} {Cambridge},\ \bibinfo
  {year} {1997})\BibitemShut {NoStop}%
\bibitem [{\citenamefont {Sakai}\ \emph {et~al.}(2016)\citenamefont {Sakai},
  \citenamefont {Lucas}, \citenamefont {Gegenwart}, \citenamefont {Stockert},
  \citenamefont {v.~L\"ohneysen},\ and\ \citenamefont {Fritsch}}]{akito16}%
  \BibitemOpen
  \bibfield  {author} {\bibinfo {author} {\bibfnamefont {A.}~\bibnamefont
  {Sakai}}, \bibinfo {author} {\bibfnamefont {S.}~\bibnamefont {Lucas}},
  \bibinfo {author} {\bibfnamefont {P.}~\bibnamefont {Gegenwart}}, \bibinfo
  {author} {\bibfnamefont {O.}~\bibnamefont {Stockert}}, \bibinfo {author}
  {\bibfnamefont {H.}~\bibnamefont {v.~L\"ohneysen}}, \ and\ \bibinfo {author}
  {\bibfnamefont {V.}~\bibnamefont {Fritsch}},\ }\href {\doibase
  10.1103/PhysRevB.94.220405} {\bibfield  {journal} {\bibinfo  {journal} {Phys.
  Rev. B}\ }\textbf {\bibinfo {volume} {94}},\ \bibinfo {pages} {220405}
  (\bibinfo {year} {2016})}\BibitemShut {NoStop}%
\bibitem [{\citenamefont {Nakatsuji}\ \emph {et~al.}(2006)\citenamefont
  {Nakatsuji}, \citenamefont {Machida}, \citenamefont {Maeno}, \citenamefont
  {Tayama}, \citenamefont {Sakakibara}, \citenamefont {Duijn}, \citenamefont
  {Balicas}, \citenamefont {Millican}, \citenamefont {Macaluso},\ and\
  \citenamefont {Chan}}]{nakatsuji06}%
  \BibitemOpen
  \bibfield  {author} {\bibinfo {author} {\bibfnamefont {S.}~\bibnamefont
  {Nakatsuji}}, \bibinfo {author} {\bibfnamefont {Y.}~\bibnamefont {Machida}},
  \bibinfo {author} {\bibfnamefont {Y.}~\bibnamefont {Maeno}}, \bibinfo
  {author} {\bibfnamefont {T.}~\bibnamefont {Tayama}}, \bibinfo {author}
  {\bibfnamefont {T.}~\bibnamefont {Sakakibara}}, \bibinfo {author}
  {\bibfnamefont {J.~v.}\ \bibnamefont {Duijn}}, \bibinfo {author}
  {\bibfnamefont {L.}~\bibnamefont {Balicas}}, \bibinfo {author} {\bibfnamefont
  {J.~N.}\ \bibnamefont {Millican}}, \bibinfo {author} {\bibfnamefont {R.~T.}\
  \bibnamefont {Macaluso}}, \ and\ \bibinfo {author} {\bibfnamefont {J.~Y.}\
  \bibnamefont {Chan}},\ }\href {\doibase 10.1103/PhysRevLett.96.087204}
  {\bibfield  {journal} {\bibinfo  {journal} {Phys. Rev. Lett.}\ }\textbf
  {\bibinfo {volume} {96}},\ \bibinfo {pages} {087204} (\bibinfo {year}
  {2006})}\BibitemShut {NoStop}%
\bibitem [{\citenamefont {Kim}\ \emph {et~al.}(2008)\citenamefont {Kim},
  \citenamefont {Bennett},\ and\ \citenamefont {Aronson}}]{kim_frustkondo08}%
  \BibitemOpen
  \bibfield  {author} {\bibinfo {author} {\bibfnamefont {M.~S.}\ \bibnamefont
  {Kim}}, \bibinfo {author} {\bibfnamefont {M.~C.}\ \bibnamefont {Bennett}}, \
  and\ \bibinfo {author} {\bibfnamefont {M.~C.}\ \bibnamefont {Aronson}},\
  }\href {\doibase 10.1103/PhysRevB.77.144425} {\bibfield  {journal} {\bibinfo
  {journal} {Phys. Rev. B}\ }\textbf {\bibinfo {volume} {77}},\ \bibinfo
  {pages} {144425} (\bibinfo {year} {2008})}\BibitemShut {NoStop}%
\bibitem [{\citenamefont {Sengupta}\ \emph {et~al.}(2010)\citenamefont
  {Sengupta}, \citenamefont {Forthaus}, \citenamefont {Kubo}, \citenamefont
  {Katoh}, \citenamefont {Umeo}, \citenamefont {Takabatake},\ and\
  \citenamefont {Abd-Elmeguid}}]{sengupta_frustkondo10}%
  \BibitemOpen
  \bibfield  {author} {\bibinfo {author} {\bibfnamefont {K.}~\bibnamefont
  {Sengupta}}, \bibinfo {author} {\bibfnamefont {M.~K.}\ \bibnamefont
  {Forthaus}}, \bibinfo {author} {\bibfnamefont {H.}~\bibnamefont {Kubo}},
  \bibinfo {author} {\bibfnamefont {K.}~\bibnamefont {Katoh}}, \bibinfo
  {author} {\bibfnamefont {K.}~\bibnamefont {Umeo}}, \bibinfo {author}
  {\bibfnamefont {T.}~\bibnamefont {Takabatake}}, \ and\ \bibinfo {author}
  {\bibfnamefont {M.~M.}\ \bibnamefont {Abd-Elmeguid}},\ }\href {\doibase
  10.1103/PhysRevB.81.125129} {\bibfield  {journal} {\bibinfo  {journal} {Phys.
  Rev. B}\ }\textbf {\bibinfo {volume} {81}},\ \bibinfo {pages} {125129}
  (\bibinfo {year} {2010})}\BibitemShut {NoStop}%
\bibitem [{\citenamefont {Kato}\ \emph {et~al.}(2008)\citenamefont {Kato},
  \citenamefont {Kosaka}, \citenamefont {Nowatari}, \citenamefont {Saiga},
  \citenamefont {Yamada}, \citenamefont {Kobiyama}, \citenamefont {Katano},
  \citenamefont {Ohoyama}, \citenamefont {Suzuki}, \citenamefont {Aso},\ and\
  \citenamefont {Iwasa}}]{kato_frustkondo08}%
  \BibitemOpen
  \bibfield  {author} {\bibinfo {author} {\bibfnamefont {Y.}~\bibnamefont
  {Kato}}, \bibinfo {author} {\bibfnamefont {M.}~\bibnamefont {Kosaka}},
  \bibinfo {author} {\bibfnamefont {H.}~\bibnamefont {Nowatari}}, \bibinfo
  {author} {\bibfnamefont {Y.}~\bibnamefont {Saiga}}, \bibinfo {author}
  {\bibfnamefont {A.}~\bibnamefont {Yamada}}, \bibinfo {author} {\bibfnamefont
  {T.}~\bibnamefont {Kobiyama}}, \bibinfo {author} {\bibfnamefont
  {S.}~\bibnamefont {Katano}}, \bibinfo {author} {\bibfnamefont
  {K.}~\bibnamefont {Ohoyama}}, \bibinfo {author} {\bibfnamefont {H.~S.}\
  \bibnamefont {Suzuki}}, \bibinfo {author} {\bibfnamefont {N.}~\bibnamefont
  {Aso}}, \ and\ \bibinfo {author} {\bibfnamefont {K.}~\bibnamefont {Iwasa}},\
  }\href {\doibase 10.1143/JPSJ.77.053701} {\bibfield  {journal} {\bibinfo
  {journal} {J. Phys. Soc. Jpn}\ }\textbf {\bibinfo {volume} {77}},\ \bibinfo
  {pages} {053701} (\bibinfo {year} {2008})}\BibitemShut {NoStop}%
\bibitem [{\citenamefont {Bercx}\ \emph {et~al.}(2017)\citenamefont {Bercx},
  \citenamefont {Goth}, \citenamefont {Hofmann},\ and\ \citenamefont
  {Assaad}}]{ALF_v1}%
  \BibitemOpen
  \bibfield  {author} {\bibinfo {author} {\bibfnamefont {M.}~\bibnamefont
  {Bercx}}, \bibinfo {author} {\bibfnamefont {F.}~\bibnamefont {Goth}},
  \bibinfo {author} {\bibfnamefont {J.~S.}\ \bibnamefont {Hofmann}}, \ and\
  \bibinfo {author} {\bibfnamefont {F.~F.}\ \bibnamefont {Assaad}},\ }\href
  {\doibase 10.21468/SciPostPhys.3.2.013} {\bibfield  {journal} {\bibinfo
  {journal} {SciPost Phys.}\ }\textbf {\bibinfo {volume} {3}},\ \bibinfo
  {pages} {013} (\bibinfo {year} {2017})}\BibitemShut {NoStop}%
\bibitem [{Note1()}]{Note1}%
  \BibitemOpen
  \bibinfo {note} {We have checked that no instabilities occur in the
  ferrromagnetic channel: $ \protect \mathaccentV {hat}05E{O}_{\protect \mathbf
  r}^{\alpha } = \protect \mathaccentV {hat}05E{S}^{tot,\alpha }_{\protect
  \mathbf { r},A} + \protect \mathaccentV {hat}05E{S}^{tot,\alpha }_{\protect
  \mathbf {r},B} $}\BibitemShut {NoStop}%
\bibitem [{\citenamefont {Binder}(1981)}]{Binder1981}%
  \BibitemOpen
  \bibfield  {author} {\bibinfo {author} {\bibfnamefont {K.}~\bibnamefont
  {Binder}},\ }\href {\doibase 10.1007/BF01293604} {\bibfield  {journal}
  {\bibinfo  {journal} {Z. Phys. B Con. Mat.}\ }\textbf {\bibinfo {volume}
  {43}},\ \bibinfo {pages} {119} (\bibinfo {year} {1981})}\BibitemShut
  {NoStop}%
\bibitem [{\citenamefont {Pujari}\ \emph {et~al.}(2016)\citenamefont {Pujari},
  \citenamefont {Lang}, \citenamefont {Murthy},\ and\ \citenamefont
  {Kaul}}]{Pujari16}%
  \BibitemOpen
  \bibfield  {author} {\bibinfo {author} {\bibfnamefont {S.}~\bibnamefont
  {Pujari}}, \bibinfo {author} {\bibfnamefont {T.~C.}\ \bibnamefont {Lang}},
  \bibinfo {author} {\bibfnamefont {G.}~\bibnamefont {Murthy}}, \ and\ \bibinfo
  {author} {\bibfnamefont {R.~K.}\ \bibnamefont {Kaul}},\ }\href {\doibase
  10.1103/PhysRevLett.117.086404} {\bibfield  {journal} {\bibinfo  {journal}
  {Phys. Rev. Lett.}\ }\textbf {\bibinfo {volume} {117}},\ \bibinfo {pages}
  {086404} (\bibinfo {year} {2016})}\BibitemShut {NoStop}%
\bibitem [{\citenamefont {Assaad}\ and\ \citenamefont
  {Imada}(1996)}]{Assaad96a}%
  \BibitemOpen
  \bibfield  {author} {\bibinfo {author} {\bibfnamefont {F.~F.}\ \bibnamefont
  {Assaad}}\ and\ \bibinfo {author} {\bibfnamefont {M.}~\bibnamefont {Imada}},\
  }\href@noop {} {\bibfield  {journal} {\bibinfo  {journal} {J. Phys. Soc.
  Jpn.}\ }\textbf {\bibinfo {volume} {65}},\ \bibinfo {pages} {189} (\bibinfo
  {year} {1996})}\BibitemShut {NoStop}%
\bibitem [{Note2()}]{Note2}%
  \BibitemOpen
  \bibinfo {note} {Due to the normalization condition, $|\protect \boldsymbol
  {\alpha }_n| = 1 $, finite values of $M_m$ lead to a reduction of the Kondo
  screening.}\BibitemShut {Stop}%
\bibitem [{\citenamefont {Tsunetsugu}\ \emph {et~al.}(1997)\citenamefont
  {Tsunetsugu}, \citenamefont {Sigrist},\ and\ \citenamefont
  {Ueda}}]{Tsunetsugu97_rev}%
  \BibitemOpen
  \bibfield  {author} {\bibinfo {author} {\bibfnamefont {H.}~\bibnamefont
  {Tsunetsugu}}, \bibinfo {author} {\bibfnamefont {M.}~\bibnamefont {Sigrist}},
  \ and\ \bibinfo {author} {\bibfnamefont {K.}~\bibnamefont {Ueda}},\
  }\href@noop {} {\bibfield  {journal} {\bibinfo  {journal} {Rev. Mod. Phys.}\
  }\textbf {\bibinfo {volume} {69}},\ \bibinfo {pages} {809} (\bibinfo {year}
  {1997})}\BibitemShut {NoStop}%
\bibitem [{\citenamefont {{J\"ulich Supercomputing Centre}}(2016)}]{Jureca16}%
  \BibitemOpen
  \bibfield  {author} {\bibinfo {author} {\bibnamefont {{J\"ulich
  Supercomputing Centre}}},\ }\href {http://dx.doi.org/10.17815/jlsrf-2-121}
  {\bibfield  {journal} {\bibinfo  {journal} {Journal of large-scale research
  facilities}\ }\textbf {\bibinfo {volume} {2}},\ \bibinfo {pages} {A62}
  (\bibinfo {year} {2016})}\BibitemShut {NoStop}%
\bibitem [{\citenamefont {Moessner}\ and\ \citenamefont
  {Sondhi}(2001{\natexlab{b}})}]{Moessner01b}%
  \BibitemOpen
  \bibfield  {author} {\bibinfo {author} {\bibfnamefont {R.}~\bibnamefont
  {Moessner}}\ and\ \bibinfo {author} {\bibfnamefont {S.~L.}\ \bibnamefont
  {Sondhi}},\ }\href {\doibase 10.1103/PhysRevB.63.224401} {\bibfield
  {journal} {\bibinfo  {journal} {Phys. Rev. B}\ }\textbf {\bibinfo {volume}
  {63}},\ \bibinfo {pages} {224401} (\bibinfo {year}
  {2001}{\natexlab{b}})}\BibitemShut {NoStop}%
\end{thebibliography}
\end{document}